\documentclass[%
 reprint,
 amsmath,amssymb,
 aps,
]{revtex4-2}

\usepackage{graphicx}
\usepackage{dcolumn}
\usepackage{bm}

\usepackage{color}
\newcommand{\ikeda}[1]{\textcolor{black}{#1}}
\newcommand{\arm}[1]{\textcolor{black}{#1}}
\newcommand{\inada}[1]{\textcolor{black}{#1}}

\begin{document}

\preprint{APS/123-QED}

\title{\ikeda{Systematic Biases from ARM-Based Sensitivity and Imaging in \arm{Compton Cameras}}}

\author{Tomonori Ikeda}
\email{tomonori.ikeda@aist.go.jp}
\affiliation{
National Institute of Advanced Industrial Science and Technology (AIST), National Metrology Institute of Japan (NMIJ), 1-1-1 Central 3, Umezono, Tsukuba, Ibaraki, 305-8563, Japan
}

\author{Tatsuya Sawano}
\affiliation{
 Graduate School of Natural Science and Technology, Kanazawa University, Kakuma, Kanazawa, 920-1192, Ishikawa, Japan
}

\author{Naomi Tsuji}
\affiliation{
Institute for Cosmic Ray Research, the university of Tokyo, Kashiwanoha 5-1-5, Kashiwa, Japan
}

\author{Yoshitaka Mizumura}
\affiliation{
 Institute of Space and Astronautical Science, Japan Aerospace Exploration Agency, Yoshinodai 3-1-1, Chuou, Sagamihara, 252-5210, Kanagawa, Japan
}



\date{\today}

\begin{abstract}

The angular resolution measure (ARM) is widely used in Compton camera analyses to characterize angular uncertainty and to define event selection criteria. However, \arm{the ARM distribution} is fundamentally different from the point spread function (PSF), which represents the true spatial imaging response of the system on the celestial sphere. 
\arm{In this study, motivated by applications in gamma-ray astrophysics, we investigated the consequences of using ARM-based definitions in sensitivity estimation and image reconstruction.}
Through Monte Carlo simulations of 662~keV gamma-rays, we demonstrated that the background rejection power of ARM-based region-of-interest selections \arm{with the ARM resolution of 7.0$^{\circ}$} is equivalent to that of a PSF resolution of 33$^\circ$. As a result, treating the ARM resolution as a surrogate for \arm{the PSF resolution} leads to significant overestimation of the signal-to-noise ratio, particularly in analyses assuming isotropic backgrounds.
We further show that image reconstructions at pixel scales finer than \arm{the PSF resolution} introduce substantial statistical uncertainties, reducing the effective number of independent spatial degrees of freedom.
These findings underscore the necessity of PSF-based analysis for accurate sensitivity evaluation and statistically robust imaging in Compton cameras, particularly in applications where the background discrimination is critical.

\end{abstract}

\maketitle


\section{Introduction}
Compton cameras are widely used for gamma-ray imaging in astrophysics~\cite{Schoenfelder_1993, Boggs_2001, e-ASTROGAM_2018, AMEGO_2021, ARAMAKI_2020, TIGER_1995}, nuclear medicine~\cite{Takahashi_2012, KABUKI2010}, and radiation monitoring~\cite{KATAOKA_2013,Kuriyama_2022, Tomono2017, KAGAYA_2015}. A central performance metric is the angular resolution measure (ARM), defined as the angular difference between the true Compton scatter angle and that reconstructed from the measured interaction kinematics. ARM is often used not only to assess angular resolution but also to define event selection criteria for improving the signal-to-noise ratio (S/N) in sensitivity studies~\cite{Kierans_2020}.
Despite its utility, the ARM resolution differs fundamentally from \arm{the resolution of the point spread function (PSF)}, which describes the spatial response of an imaging system to a point source on the celestial sphere. \ikeda{In conventional linear imaging systems based on geometrical optics, such as optical and X-ray telescopes, the PSF resolution is defined as the minimum angular separation at which the brightness (or radiance) of two sources can be independently preserved. Because such systems preserve brightness---a fundamental physical quantity---the PSF governs the limit of meaningful spatial resolution and allows for direct, quantitative analysis of reconstructed images.}

The point-source sensitivity of a gamma-ray telescope, defined as the minimum detectable flux $F_{\mathrm{min}}$, is typically expressed as
\begin{equation}
    F_{\mathrm{min}} = \frac{n\sqrt{N_{\mathrm{S}}+N_{\mathrm{B}}}}{A_{\mathrm{eff}}T}
\end{equation}
where, $n$ is the detection significance, $N_{\mathrm{S}}$ and $N_{\mathrm{B}}$ are the numbers of source and background photons, $A_{\mathrm{eff}}$ is the effective area, and $T$ is the observation time.
In this framework, the background component $N_{\mathrm{B}}$ must be evaluated over a spatial region defined by the PSF. For isotropic backgrounds—such as the extragalactic diffuse gamma-ray—this is straightforward in celestial coordinates. 
However, using \arm{the ARM distribution} as a proxy for PSF introduces a mismatch between coordinate systems, complicating the evaluation of background counts. This often leads to systematic underestimation of the background and, consequently, overestimation of sensitivity.

Several studies have treated the ARM resolution as a substitute for the PSF resolution, or used it in image reconstruction methods such as the Maximum Entropy Method~\cite{Strong1995} and other statistical approaches~\cite{Knodlseder_1999, IKEDA201446, Wilderman_1998}. Images are often reconstructed at resolutions finer than the intrinsic PSF \arm{resolution}, which introduces strong statistical artifacts and uncertainties. 
Moreover, reconstructed Compton images do not represent direct measurements of photon counts. Each pixel value corresponds to a statistical estimator—typically the posterior probability density of the source intensity—rather than a directly observed quantity.
This differs significantly from \ikeda{general imaging systems based on geometrical optics}, where image pixels reflect measured photon counts and represent physically meaningful observables. As such, Compton images reconstructed via statistical algorithms should not be referred to or interpreted in the same way as direct observation maps.

Furthermore, most studies on Compton cameras present \arm{the ARM resolution} as a surrogate for angular resolution, or omit any presentation of \arm{the PSF resolution}~\cite{Kierans_2022_Springer, ARAMAKI_2020}, which can be misleading.
ARM is a local angular uncertainty parameter rather than a measure of imaging performance. Presenting it as angular resolution can mislead readers and result in systematic underestimation of background, overinterpretation of reconstructed features, and mischaracterization of the instrument’s capabilities. 
We argue that the PSF—not the ARM—should be considered the fundamental descriptor of angular resolution in Compton imaging systems. This allows for consistent performance comparisons across different wavelength regimes and instruments.

In this work, we quantitatively investigate the consequences of using ARM-based definitions for sensitivity evaluation and image reconstruction. We show that such practices lead to systematic biases and degraded statistical reliability. In particular, we examine the effects of using pixel scales finer than the PSF resolution and analyze the resulting statistical structure of the reconstructed images.
Sec.~\ref{sec:param} defines the relevant parameters in Compton cameras and evaluates the PSF in celestial coordinates. 
Sec.~\ref{sec:bg} presents background rejection performance in the Compton data space and compares it with the PSF-based imaging selection. 
Section~\ref{sec:imaging} analyzes the statistical impact of image reconstruction at ARM-scale pixel resolutions and discusses its limitations.

\section{Definition of Parameters and Point Spread Function on the Celestial Sphere\label{sec:param}}
In this section, we define the key physical and geometrical parameters relevant to Compton imaging, and clarify the distinction between the angular resolution measure (ARM) and the point spread function (PSF). 

\begin{figure}
\includegraphics[width=8.5cm]{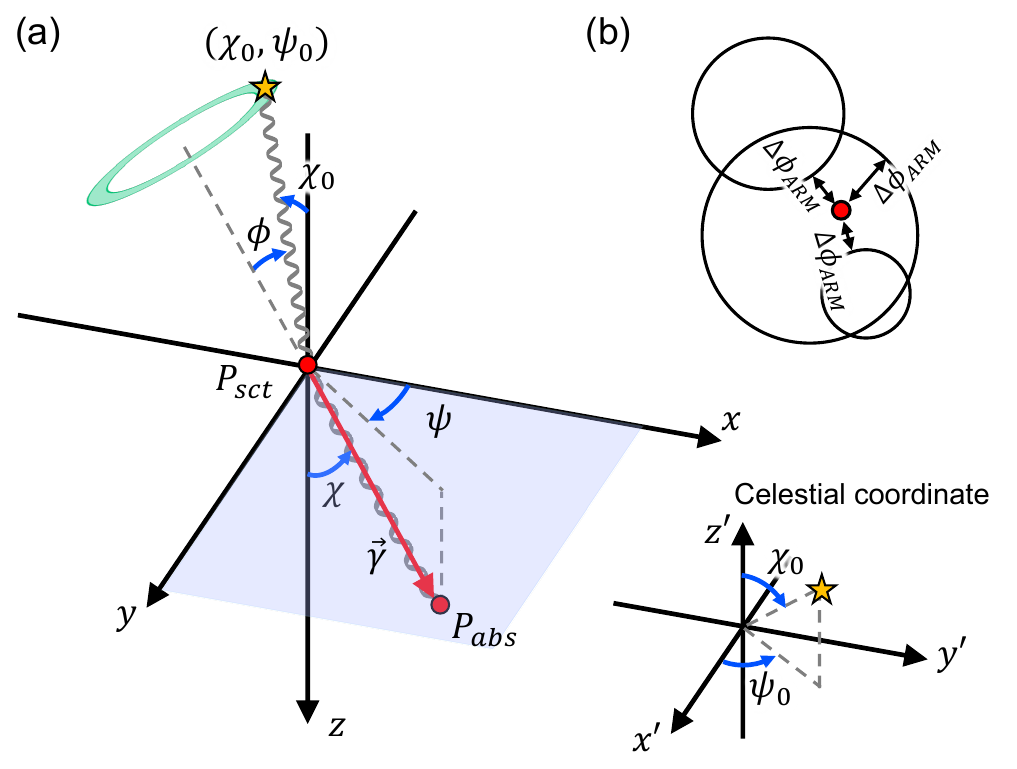}
\caption{\label{fig:geometry} 
(a) Geometry of a Compton scattering event. A gamma-ray from sky direction \((\chi_0, \psi_0)\) undergoes scattering at \(P_{\mathrm{sct}}\), and the scattered photon is absorbed at \(P_{\mathrm{abs}}\). The measured direction \(\vec{\gamma}\) and \arm{the true incident gamma-ray direction} define the scattering angle \(\phi\).
\ikeda{The blue area indicates the $x$-$y$ plane.}
\arm{The $z$-axis of the detector frame is aligned with the $z’$-axis of the celestial coordinate system, but their directions are opposite.}
\inada{(b) Schematic illustration of the angular resolution measure. The ARM  is defined as the smallest angular distance between the source location and each event circle.}}
\end{figure}

Fig.~\ref{fig:geometry} illustrates the coordinate system and angular definitions of Compton scattering events. 
A gamma-ray photon from the sky direction $(\chi_{0}, \psi_0)$, where $\chi_0$ and $\psi_0$ denote the polar and azimuthal angles in celestial coordinates, undergoes a Compton interaction at the point $P_{\mathrm{sct}}$. 
The scattered photon subsequently travels along the direction $\vec{\gamma}$ and is fully absorbed at the point $P_{\mathrm{abs}}$.
In the detector coordinate, the scattered photon trajectory is parameterized by the angles $\chi$ and $\psi$, which represent the polar and azimuthal angles, respectively. 
\arm{The $z$-axis of the detector coordinate system is aligned with the $z'$-axis of the celestial coordinate system, but points in the opposite direction.}
The plane containing both the incident and scattered photon directions defines the Compton scattering plane. The angle $\phi$ is the kinematic scattering angle between the incident direction and $\vec{\gamma}$, and it is constrained by the Compton scattering formula:
\begin{equation}\label{eq:compton}
    \cos\phi = 1- \frac{m_\mathrm{e}c^{2}}{E_{\gamma}+K_{\mathrm{e}}}\frac{K_{\mathrm{e}}}{E_{\gamma}},
\end{equation}
where $E_{\gamma}$ and $K_{\mathrm{e}}$ are the energies of the scattered photon and Compton electron, respectively, and $m_\mathrm{e} c^2$ is the electron rest energy.

The ARM is defined as the difference between the measured scattering angle and the expected one:
\begin{equation}
\arm{\Delta}\phi_{\mathrm{ARM}}=\arccos(\vec{s}\cdot\vec{\gamma})-\phi,
\end{equation}
where $\vec{s}$ is the reconstructed incident photon direction, and $\phi$ is calculated via Eq.~\eqref{eq:compton}.
\arm{Fig.~\ref{fig:geometry}(b) shows three event circles and the ARM corresponds to the smallest angular distance between the source location and each event circle.}

\begin{figure}[h]
\includegraphics[width=8.5cm]{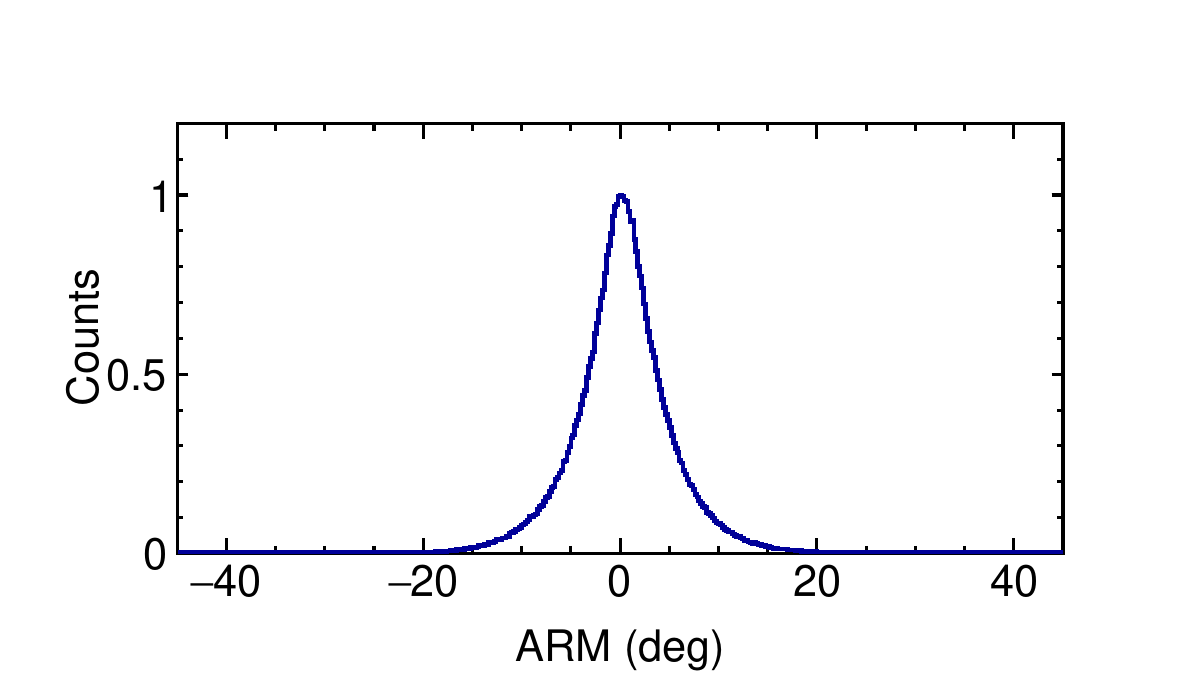}
\caption{\label{fig:arm} 
ARM distribution from Monte Carlo simulation with \ikeda{662~keV} photons, 10\% energy resolution, and 1$^{\circ}$ angular resolution for the scattered photon. The FWHM is approximately \ikeda{7.0$^{\circ}$}.}
\end{figure}

The ARM distribution is affected by several factors, including Doppler broadening (due to the binding of electrons in atomic orbitals), finite energy resolution, and limited position resolution of the detector. These factors smear the reconstructed angle $\phi$, degrading the ARM \arm{distribution}.
\arm{In particular, Doppler broadening imposes a fundamental physical limitation on the ARM and depends on the atomic number of the detector material~\cite{Zoglauer_thesis_2005}. In this study, we assume detector material independence and therefore neglect the effect of Doppler broadening.}
Fig.~\ref{fig:arm} shows the ARM distribution obtained by Monte Carlo simulation for \ikeda{662~keV} incident gamma rays, assuming an energy resolution of 10\% and a 1$^{\circ}$ angular resolution for the scattered photon. To reflect the actual detector geometry, only events with a Compton scattering angle $\phi < 90^{\circ}$ were included.
Under these conditions, the ARM resolution—defined as the full width at half maximum (FWHM)—is approximately \ikeda{$7.0^{\circ}$}.

\begin{figure}[h]
\includegraphics[width=8.5cm]{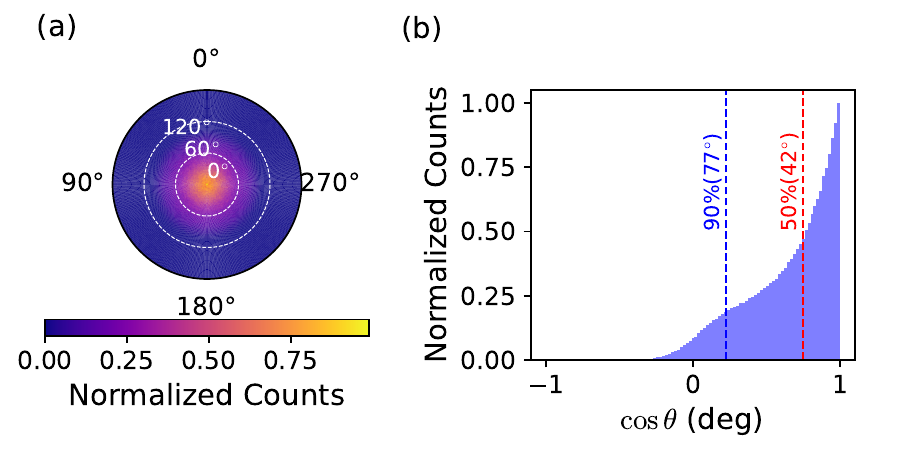}
\caption{\label{fig:psf} 
(a) Simulated sky image of a point source at \((\chi_0, \psi_0) = (0^\circ, 0^\circ)\), \arm{shown in the Lambert equal-area conic projection covering the entire celestial sphere ($4\pi$ steradians).} (b) Distribution of \(\cos\theta\) between the true and reconstructed directions. The 50\% containment radius defines the PSF resolution as \ikeda{42$^{\circ}$}.}
\end{figure}

While ARM characterizes the angular reconstruction accuracy of the Compton scattering angle, it does not describe how reconstructed directions distribute spatially on the sky.
In conventional Compton imaging without electron tracking, the incident gamma-ray direction cannot be uniquely determined and is instead reconstructed as lying somewhere along the Compton circle, as illustrated in Fig.~\ref{fig:geometry}.
To evaluate the PSF, we projected the reconstructed direction onto the celestial sphere. 
In this study, we simulate a point source at $(\chi_0, \psi_0) = (0^\circ, 0^\circ)$ under the same detector conditions as above. 
Fig.~\ref{fig:psf}(a) shows the resulting sky map \arm{in the Lambert equal-area conic projection}, and Fig.~\ref{fig:psf}(b) shows the distribution of $\cos\theta$, where $\theta$ is the angular separation between the true and reconstructed directions. 
The PSF resolution is defined as the angular radius enclosing 50\% of the reconstructed photons, yielding a value of approximately \ikeda{42$^{\circ}$}, significantly broader than the ARM resolution.
This difference clearly demonstrates that the ARM resolution and the PSF resolution are fundamentally distinct quantities.

\inada{Again, the ARM resolution is a quality metric derived solely from the smallest angular distance between the source location and each event circle, as illustrated in Fig.~\ref{fig:geometry}(b). In contrast, the PSF resolution reflects the overall imaging response by considering the entire set of event circles. Since conventional Compton cameras lack information about the recoil electron direction, they cannot constrain the incident photon to a specific segment of the Compton circle. As a result, the full area of each Compton circle must be taken into account in the PSF evaluation.}

\section{Background Evaluation Using ARM\label{sec:bg}}
In this section, we evaluate the effectiveness and limitations of using ARM-based event selection for background suppression. Although \arm{the ARM-based selection} are widely used in Compton camera analyses, their geometric interpretation in sky coordinates is nontrivial. We demonstrate that background rejection based solely on ARM does not yield equivalent results to PSF-based selection and may lead to systematic underestimation of the background contribution.

\ikeda{To provide context, we note that the representation of Compton events in data space has been a critical tool since the early days of gamma-ray Compton imaging. In particular, the Compton Data Space (CDS) was introduced in the COMPTEL mission as a three-dimensional space in which each event is mapped by its reconstructed Compton scattering angle and direction~\cite{Schoenfelder_1993}. The CDS provides a convenient framework for analyzing both signal and background distributions in a unified geometric context, and it remains widely used in modern analyses.}

\begin{figure}
\includegraphics[width=8.5cm]{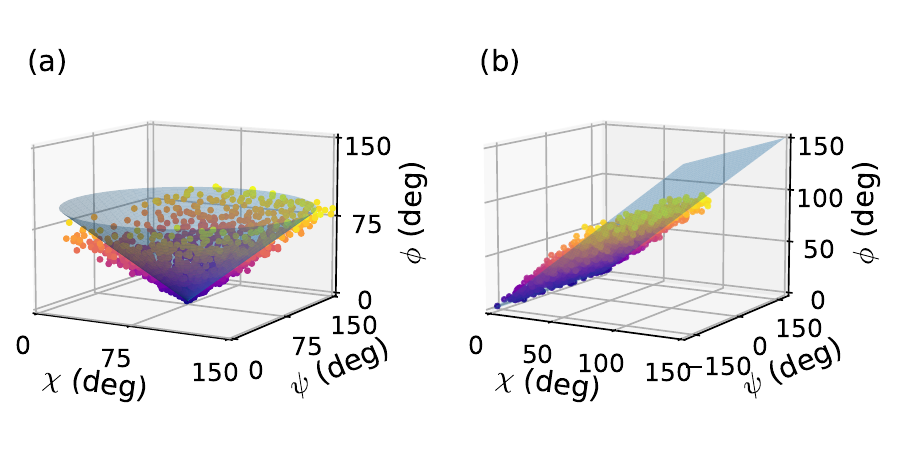}
\caption{\label{fig:cds} 
Compton data space (CDS) for simulated point sources. The color scale indicates the Compton scattering angle $\phi$. (a) \ikeda{Events from an off-axis source located at \((75^\circ, 75^\circ)\), forming a hollow conical structure.}   (b) For an on-axis source at \((0^\circ, 0^\circ)\), the events align along a diagonal plane.}
\end{figure}

Fig.~\ref{fig:cds} illustrates the CDS, which encodes the relationship among the measured Compton scattering angle $\phi$ and the reconstructed photon direction $(\chi, \psi)$ in detector coordinates. Each detected event is represented as a point in this three-dimensional space.
Fig.~\ref{fig:cds}(a) shows simulated events from a point source located at $(\chi_0, \psi_0) = (75^\circ, 75^\circ)$.
The events form a hollow conical structure, where the apex of the cone corresponds to the source position. The blue surface represents the ideal conical geometry defined by the Compton kinematics. Since \(\chi\) and \(\psi\) increase at the same rate in this configuration, the opening angle of the cone is 90$^\circ$.
In contrast, Fig.~\ref{fig:cds}(b) shows the CDS for a point source located at the origin $(0^\circ, 0^\circ)$, which forms a diagonal plane in the CDS due to the symmetry of the on-axis configuration.

\begin{figure}[h]
\includegraphics[width=8.5cm]{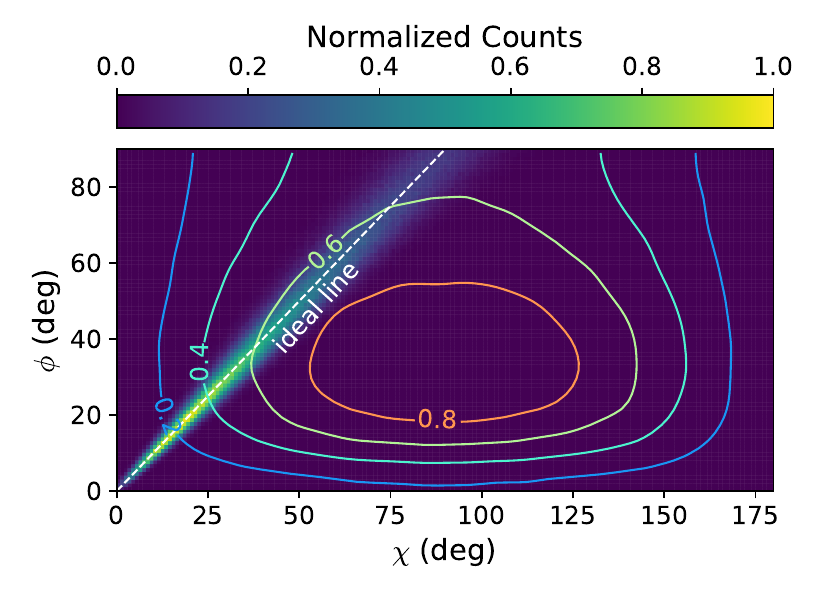}
\caption{\label{fig:cds_cut_a} 
\ikeda{Two-dimensional histogram of Compton data space (CDS) projected onto the $(\chi, \phi)$ plane for an on-axis source. The color scale indicates the normalized event density. The diagonal dashed line labeled “ideal line” corresponds to $\phi = \chi$, along which signal events are expected to align according to Compton kinematics. The contour lines represent levels of constant probability for isotropic background events.} 
}
\end{figure}

To simplify the analysis, we project the three-dimensional CDS onto a two-dimensional $(\chi, \phi)$ plane, hereafter referred to as the 2D-CDS, by integrating over the $\psi$ direction. Fig.~\ref{fig:cds_cut_a} shows the resulting 2D histogram projection for the on-axis source. Signal events align closely along the diagonal line $\phi = \chi$, which we refer to as the “ideal line.” The ARM value for each event corresponds to the perpendicular distance from this line.

\begin{figure}[h]
\includegraphics[width=8.5cm]{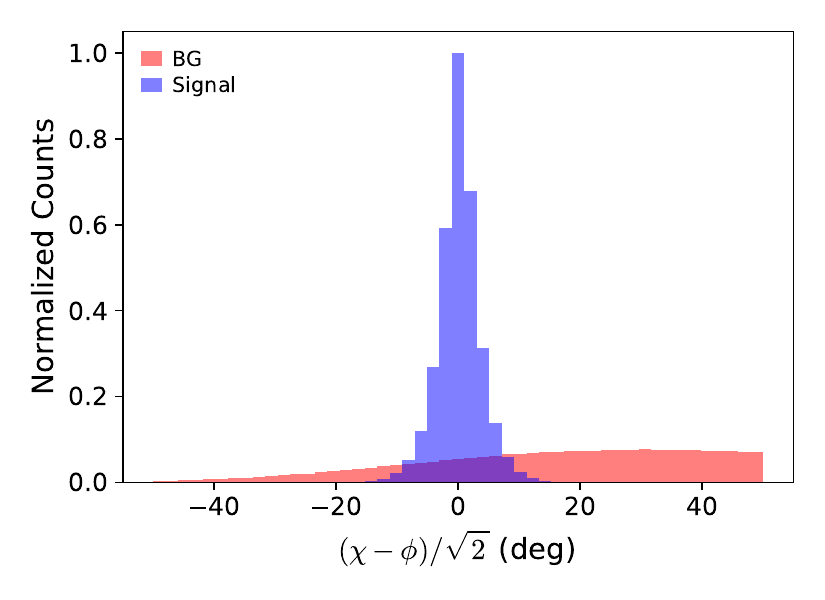}
\caption{\label{fig:cds_cut_b} 
Distribution of perpendicular distances from the ideal line for signal events (blue) and isotropic background events (red). The ideal line corresponds to the expected relation between $\chi$ and $\phi$ in Compton kinematics.}
\end{figure}

Fig.~\ref{fig:cds_cut_b} shows the distribution of this distance for signal events.
Thus, the ARM-based selection—such as selecting events within the FWHM around the ideal line—can effectively suppress background in this space.

However, to correctly evaluate the signal-to-noise ratio (S/N), it is essential to consider how isotropic background events are distributed in the 2D-CDS. Notably, the projection from celestial coordinates to the CDS does not preserve solid angle. As a result, an ARM-based selection does not correspond to a spatially uniform region on the sky and therefore does not reflect the true PSF-equivalent area.

To investigate this effect, we performed a Monte Carlo simulation of isotropic background events, as shown in Fig.~\ref{fig:cds_cut_a} and \ref{fig:cds_cut_b} , presented as a contour plot and a red histogram, respectively. 
\ikeda{In this simulation, we generated 1,000,000 events each for the signal and background distributions to ensure sufficient statistical accuracy.}
We defined the region of interest (ROI) in Fig.~\ref{fig:cds_cut_b} as the FWHM width of the ARM distribution, which is \ikeda{7.0$^\circ$}, and applied \arm{this selection} to both signal and background datasets. 
Assuming equal total event counts for signal and background, we obtained an S/N of \ikeda{12.2}. 
This value corresponds to the S/N that would be achieved on the celestial sphere using \arm{a PSF distribution} of \ikeda{33$^\circ$}.

This discrepancy demonstrates that \arm{the ARM resolution} is not equivalent to \arm{the PSF resolution}. Using ARM as a proxy for the PSF in sensitivity estimation leads to a systematic underestimation of background contributions and, consequently, an overestimation of instrument performance. Accurate sensitivity evaluation requires either PSF-defined ROIs in celestial coordinates or careful modeling of the background distribution within the 2D-CDS, taking into account the nonuniform mapping of isotropic backgrounds.

\section{Imaging and Statistical Limitations at Sub-PSF Scales}\label{sec:imaging}
In this section, we explore the statistical limitations associated with reconstructing gamma-ray images using pixel scales finer than the intrinsic PSF resolution of the Compton camera. 
While high-resolution imaging associated with the ARM resolution is often \ikeda{desirable}, it leads to increased uncertainty, thereby degrading the effective information content of the image.

The measurement process of Compton cameras can be described in the following way:
\begin{equation}\label{eq:measurement}
    D(\bm{d}) = R(\bm{d};\phi_{\mathrm{lon}},\theta_{\mathrm{lat}})  \times S(\phi_{\mathrm{lon}},\theta_{\mathrm{lat}}) + B(\bm{d}),
\end{equation}
where $D(\bm{d})$ represents the measured data in detection parameter space $\bm{d}$, $R(\bm{d};\phi_{\mathrm{lon}},\theta_{\mathrm{lat}})$ is the system response function, $S(\phi_{\mathrm{lon}},\theta_{\mathrm{lat}})$  is the true sky intensity, and $B(\bm{d})$ is the background contribution. In the case of Compton data, $\bm{d}$ corresponds to $(\chi, \psi, \phi)$, although only two degrees of freedom remain after projection onto celestial coordinates.

The goal of image reconstruction is to invert Eq.~(\ref{eq:measurement}) and estimate $\bm{S}$ from measured data $\bm{D}$, given a known response function and background model, where $\bm{S}$ and $\bm{D}$ are vectorized. To ensure numerical stability and mitigate the amplification of uncertainty, regularization is often introduced. In this work, we adopt Tikhonov regularization:
\begin{equation}\label{eq:tikhonov}
 \hat{\bm{S}} = \arg\min_{\bm{S}} \left( \| R\bm{S} - \bm{D} \|^2 + \lambda \| \bm{S} \|^2 \right)   
\end{equation}
where $\hat{\bm{S}}$ is the solution that minimizes the objective function, and $\lambda$ is the regularization parameter.

\ikeda{The regularization parameter $\lambda$ plays a crucial physical role: it controls the balance between fidelity to the measured data and smoothness of the reconstructed image. A small $\lambda$ allows the reconstruction to follow the data closely but risks amplifying statistical fluctuations and noise, especially in the presence of fine-scale pixelation below the PSF resolution scale. A large $\lambda$, on the other hand, suppresses noise but may overly smooth the image, reducing sensitivity to real features. Thus, $\lambda$ effectively defines the angular scale at which image features are deemed statistically significant, and its optimization reflects the trade-off between resolution and stability.}

To evaluate the effects of pixel scale, we discretize the sky into 768 equal-area pixels, corresponding to a pixel size of 4.1$^\circ$. The response matrix $R$ is thus 768~$\times$~768, and both $\bm{S}$ and $\bm{D}$ are 768-dimensional vectors. 
In the case of \arm{the PSF resolution} of 4$^\circ$ being the same as the pixel size, the response function becomes nearly diagonal.

We modeled the response function of a Compton camera assuming the same detector characteristics in Sec~\ref{sec:bg}. Again, \arm{the actual PSF resolution} is approximately \ikeda{$42^\circ$}, which is substantially broader than the pixel size. Under such conditions, regularization is necessary to mitigate statistical uncertainty and stabilize the reconstruction.

To evaluate the statistical behavior of image reconstruction, we simulate a point source at $(\phi_{\mathrm{lon}}, \theta_{\mathrm{lat}}) = (0^\circ, 0^\circ)$, generating 10,000 Poisson-distributed photons per dataset and adding 1\% background \arm{events}. We repeat this simulation 1,000 times and evaluate the variance across reconstructions.

\begin{figure}
\includegraphics[width=8.5cm]{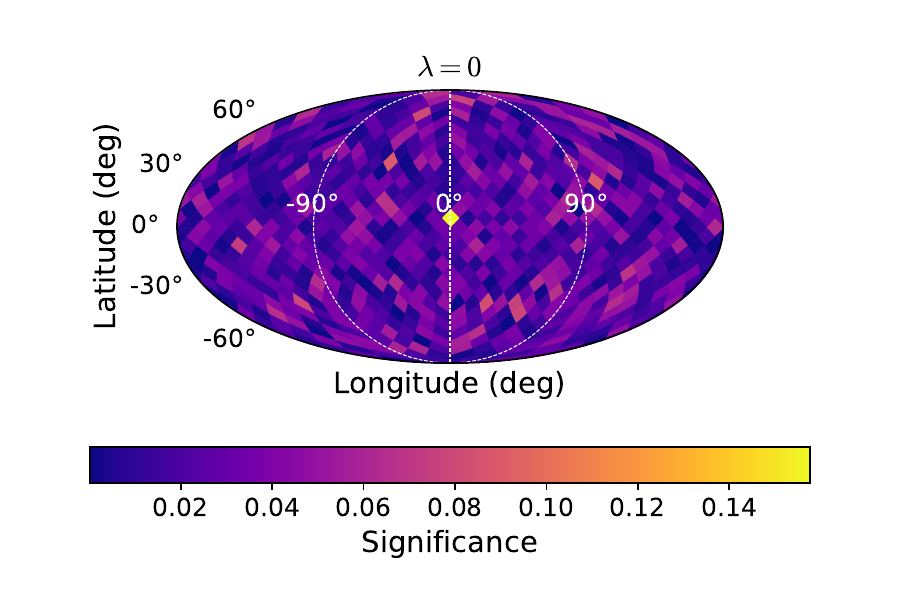}
\caption{\label{fig:std_map_a} 
Significance map of reconstructed images from 1000 data-set of a simulated point source at the center of the field of view.  
Result without regularization (\(\lambda = 0\)) shows large fluctuations due to overfitting of noise in fine pixels.  
}
\end{figure}

Fig.~\ref{fig:std_map_a} shows \ikeda{the significance maps} for reconstructions with $\lambda = 0$ (no regularization). The unregularized case exhibits large fluctuations resulting the low significance due to overfitting of statistical noise in fine-scale pixels.

To quantify reconstruction accuracy, we decompose the mean squared error (MSE) into bias and variance:
\begin{equation}
\mathrm{MSE} = \frac{1}{N}\sum _{n}(\bar{S}^{\mathrm{exp}}_{n}-S_{n})^{2} 
    + \frac{1}{NL}\sum_{n}\sum_{l}(S_{n,l}^{\mathrm{exp}}-\bar{S}^{\mathrm{exp}}_{n})^{2}.
\label{eq:mse}
\end{equation}
Here, $N$ is the total number of sky pixels, and $L$ is the number of independent reconstructed datasets. $S_n$ denotes the true intensity in pixel $n$, while $S_{n,l}^{\mathrm{exp}}$ is the reconstructed intensity in pixel $n$ for the $l$-th dataset. 
$\bar{S}^{\mathrm{exp}}_{n}$ represents the ensemble-averaged reconstructed intensity in pixel $n$, computed as $\bar{S}^{\mathrm{exp}}_{n} = \frac{1}{L}\sum_l S^{\mathrm{exp}}_{n,l}$. The first term in Eq.~(\ref{eq:mse}) corresponds to the squared bias, and the second term corresponds to the variance of the reconstruction.

\begin{figure}
\includegraphics[width=8.5cm]{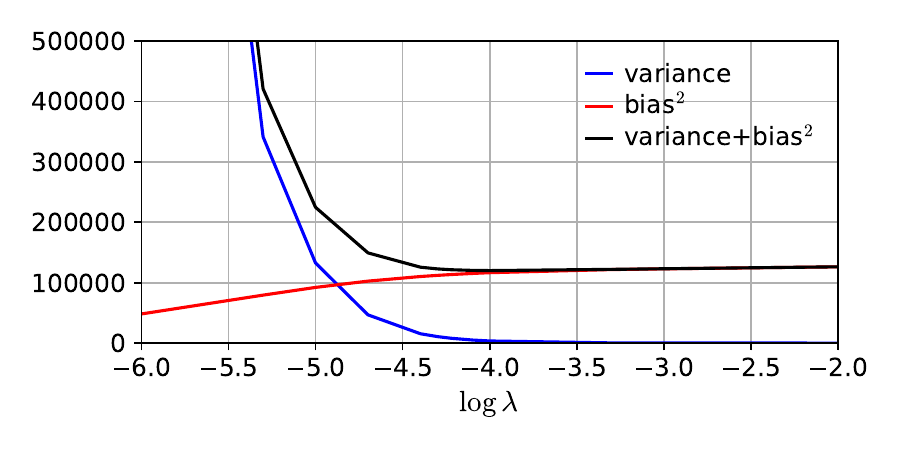}
\caption{\label{fig:std_sig} 
Bias–variance decomposition of the mean squared error (MSE) as a function of the regularization parameter \(\lambda\) in Tikhonov-regularized image reconstruction.  
The total MSE (black line) reaches a minimum where the increase in bias (red line) balances the reduction in variance (blue line).  
This optimal point reflects the trade-off between overfitting and oversmoothing, and guides the selection of \(\lambda\) for stable high-fidelity reconstruction.}
\end{figure}

\begin{figure}
\includegraphics[width=8.5cm]{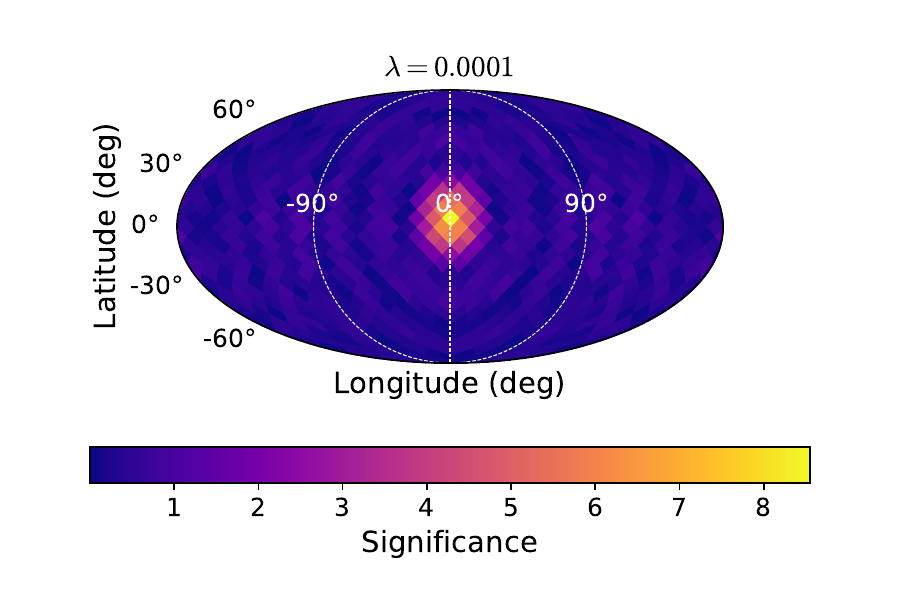}
\caption{\label{fig:std_map_b} 
Significance map of reconstructed images from 1000 data-set of a simulated point source at the center of the field of view.  A regularization term (\(\lambda = 0.0001\)) significantly suppresses pixel-to-pixel noise while preserving source localization.  
}
\end{figure}

Fig.~\ref{fig:std_sig} shows the dependence of MSE, bias$^2$, and variance on the regularization parameter $\lambda$. An optimal value of $\lambda \sim 0.0001$ minimizes the total error.
Fig.~\ref{fig:std_map_b} shows the significance map with $\lambda=0.0001$. The regularization term significantly suppresses the uncertainty while retaining source localization.

\begin{figure}
\includegraphics[width=8.5cm]{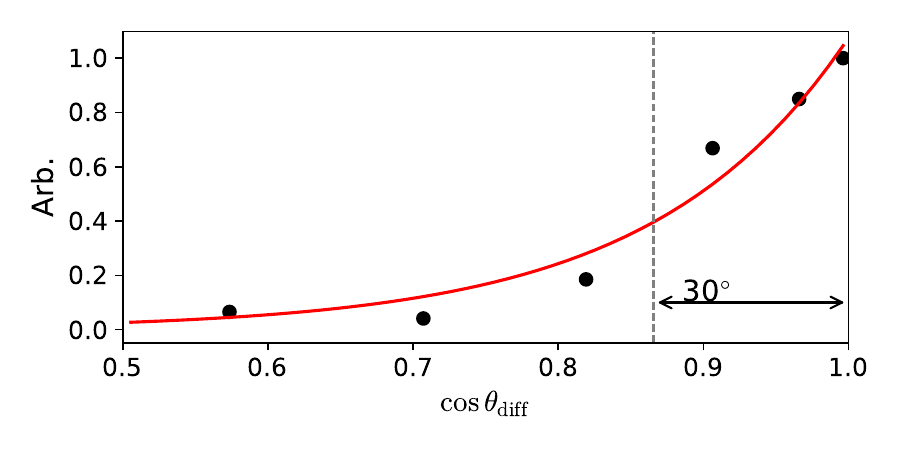}
\caption{\label{fig:length} 
Correlation function of the reconstructed image, computed as a function of the angular separation \(\theta_{\mathrm{diff}}\) between sky pixels.}
\end{figure}

We also evaluate the spatial correlation function defined as the normalized correlation between pixel $i$ and the point source pixel separated by angular distance $\theta_{\mathrm{diff}}$. As shown in Fig.~\ref{fig:length}, the correlation decays to $1/e$ at approximately \ikeda{$30^\circ$}, which we define as the correlation length. This corresponds to the angular scale over which pixels are statistically independent.

Since the PSF governs the smallest resolvable angular scale, pixel sizes much smaller than \arm{the PSF resolution} do not increase effective resolution but instead amplify statistical noise. The correlation length determines the effective number of independent degrees of freedom in the image and justifies matching pixel size to the PSF resolution for optimal imaging performance.

These results highlight a fundamental limitation: using pixel sizes much smaller than \arm{the PSF resolution} does not improve spatial resolution but instead increases noise and inter-pixel correlation. In particular, image reconstructions using ARM-scale pixels exhibit large statistical uncertainties.
Optimal pixel size should be matched to \arm{the PSF resolution} to balance resolution and statistical reliability.

\section{Conclusion and Discussion}
Our results reveal a fundamental disconnect between the commonly used ARM and the PSF, which characterizes the true imaging performance of Compton cameras. While ARM is often used as a surrogate for PSF in both sensitivity evaluation and image reconstruction, our study demonstrates that this substitution can lead to serious misinterpretations of the instrument’s capability.

In sensitivity evaluation, selecting a region of interest (ROI) based on ARM—typically a narrow angular window centered on the expected scatter angle—ignores the spatial projection of event distributions onto the celestial sphere. Our simulations show that the background suppression power of an ARM-defined ROI is effectively equivalent to that of a PSF with a broad angular radius of approximately 33$^\circ$. Consequently, if one assumes that \arm{the ARM resolution} and \arm{the PSF resolution} are equivalent and estimates the signal-to-noise ratio (S/N) under the assumption of an isotropic background, the sensitivity will be substantially overestimated.

In image reconstruction, we further demonstrate that using pixel sizes significantly smaller than \arm{the PSF resolution} does not improve spatial resolution. Instead, it introduces large statistical fluctuations. 
The number of effective degrees of freedom in the reconstructed image is limited not by the ARM resolution but by the correlation length, which we also find to be approximately 30$^{\circ}$, comparable to the PSF resolution. 
Consequently, employing pixel sizes at the ARM resolution scale serves only to amplify noise and may degrade, rather than improve, the detectability of astrophysical sources.

The only physically meaningful path to improving the PSF in Compton cameras lies in resolving the directional ambiguity of the incident gamma-ray~\cite{TANIMORI2017,SCHONFELDER2004}.
In conventional Compton cameras, attempts to enhance resolution by narrowing the acceptance of the Compton scattering angle—such as restricting to small-angle events—can reduce \arm{the PSF resolution}, but this reduces detection efficiency, often by more than an order of magnitude.
The only viable method to reduce \arm{the PSF resolution} without sacrificing efficiency is to measure the direction of the recoil electron. This additional information allows the reconstructed source direction to be constrained to an arc segment rather than the full Compton circle, thereby substantially improving angular resolution~\cite{Ikeda2021,Tanimori_2015,TANIMORI2004263,2023Ikeda}.

These findings have direct implications for both instrumentation and data analysis in Compton imaging systems. 
Treating ARM as a proxy for PSF, whether for simplicity or tradition, can lead to serious misinterpretations of instrument capability, especially in low signal-to-background regimes common in astrophysics and medical imaging.

\begin{acknowledgments}
This study was supported by the Japan Society for the Promotion of Science (JSPS) KAKENHI Grant-in-Aids for Scientific Research (Grant Numbers 22KJ1766, and 24K00643).
Furthermore, we would like to thank Toru Tanimori and Atsushi Takada, Tomohiro Inada for the insightful discussion.

\end{acknowledgments}

\appendix


\nocite{*}

\bibliography{apssamp}

\begin{thebibliography}{25}%
\makeatletter
\providecommand \@ifxundefined [1]{%
 \@ifx{#1\undefined}
}%
\providecommand \@ifnum [1]{%
 \ifnum #1\expandafter \@firstoftwo
 \else \expandafter \@secondoftwo
 \fi
}%
\providecommand \@ifx [1]{%
 \ifx #1\expandafter \@firstoftwo
 \else \expandafter \@secondoftwo
 \fi
}%
\providecommand \natexlab [1]{#1}%
\providecommand \enquote  [1]{``#1''}%
\providecommand \bibnamefont  [1]{#1}%
\providecommand \bibfnamefont [1]{#1}%
\providecommand \citenamefont [1]{#1}%
\providecommand \href@noop [0]{\@secondoftwo}%
\providecommand \href [0]{\begingroup \@sanitize@url \@href}%
\providecommand \@href[1]{\@@startlink{#1}\@@href}%
\providecommand \@@href[1]{\endgroup#1\@@endlink}%
\providecommand \@sanitize@url [0]{\catcode `\\12\catcode `\$12\catcode `\&12\catcode `\#12\catcode `\^12\catcode `\_12\catcode `\%12\relax}%
\providecommand \@@startlink[1]{}%
\providecommand \@@endlink[0]{}%
\providecommand \url  [0]{\begingroup\@sanitize@url \@url }%
\providecommand \@url [1]{\endgroup\@href {#1}{\urlprefix }}%
\providecommand \urlprefix  [0]{URL }%
\providecommand \Eprint [0]{\href }%
\providecommand \doibase [0]{https://doi.org/}%
\providecommand \selectlanguage [0]{\@gobble}%
\providecommand \bibinfo  [0]{\@secondoftwo}%
\providecommand \bibfield  [0]{\@secondoftwo}%
\providecommand \translation [1]{[#1]}%
\providecommand \BibitemOpen [0]{}%
\providecommand \bibitemStop [0]{}%
\providecommand \bibitemNoStop [0]{.\EOS\space}%
\providecommand \EOS [0]{\spacefactor3000\relax}%
\providecommand \BibitemShut  [1]{\csname bibitem#1\endcsname}%
\let\auto@bib@innerbib\@empty
\bibitem [{\citenamefont {{Schoenfelder}}\ \emph {et~al.}(1993)\citenamefont {{Schoenfelder}}, \citenamefont {{Aarts}}, \citenamefont {{Bennett}}, \citenamefont {{de Boer}}, \citenamefont {{Clear}}, \citenamefont {{Collmar}}, \citenamefont {{Connors}}, \citenamefont {{Deerenberg}}, \citenamefont {{Diehl}}, \citenamefont {{von Dordrecht}}, \citenamefont {{den Herder}}, \citenamefont {{Hermsen}}, \citenamefont {{Kippen}}, \citenamefont {{Kuiper}}, \citenamefont {{Lichti}}, \citenamefont {{Lockwood}}, \citenamefont {{Macri}}, \citenamefont {{McConnell}}, \citenamefont {{Morris}}, \citenamefont {{Much}}, \citenamefont {{Ryan}}, \citenamefont {{Simpson}}, \citenamefont {{Snelling}}, \citenamefont {{Stacy}}, \citenamefont {{Steinle}}, \citenamefont {{Strong}}, \citenamefont {{Swanenburg}}, \citenamefont {{Taylor}}, \citenamefont {{de Vries}},\ and\ \citenamefont {{Winkler}}}]{Schoenfelder_1993}%
  \BibitemOpen
  \bibfield  {author} {\bibinfo {author} {\bibfnamefont {V.}~\bibnamefont {{Schoenfelder}}}, \bibinfo {author} {\bibfnamefont {H.}~\bibnamefont {{Aarts}}}, \bibinfo {author} {\bibfnamefont {K.}~\bibnamefont {{Bennett}}}, \bibinfo {author} {\bibfnamefont {H.}~\bibnamefont {{de Boer}}}, \bibinfo {author} {\bibfnamefont {J.}~\bibnamefont {{Clear}}}, \bibinfo {author} {\bibfnamefont {W.}~\bibnamefont {{Collmar}}}, \bibinfo {author} {\bibfnamefont {A.}~\bibnamefont {{Connors}}}, \bibinfo {author} {\bibfnamefont {A.}~\bibnamefont {{Deerenberg}}}, \bibinfo {author} {\bibfnamefont {R.}~\bibnamefont {{Diehl}}}, \bibinfo {author} {\bibfnamefont {A.}~\bibnamefont {{von Dordrecht}}}, \bibinfo {author} {\bibfnamefont {J.~W.}\ \bibnamefont {{den Herder}}}, \bibinfo {author} {\bibfnamefont {W.}~\bibnamefont {{Hermsen}}}, \bibinfo {author} {\bibfnamefont {M.}~\bibnamefont {{Kippen}}}, \bibinfo {author} {\bibfnamefont {L.}~\bibnamefont {{Kuiper}}}, \bibinfo {author} {\bibfnamefont {G.}~\bibnamefont {{Lichti}}}, \bibinfo
  {author} {\bibfnamefont {J.}~\bibnamefont {{Lockwood}}}, \bibinfo {author} {\bibfnamefont {J.}~\bibnamefont {{Macri}}}, \bibinfo {author} {\bibfnamefont {M.}~\bibnamefont {{McConnell}}}, \bibinfo {author} {\bibfnamefont {D.}~\bibnamefont {{Morris}}}, \bibinfo {author} {\bibfnamefont {R.}~\bibnamefont {{Much}}}, \bibinfo {author} {\bibfnamefont {J.}~\bibnamefont {{Ryan}}}, \bibinfo {author} {\bibfnamefont {G.}~\bibnamefont {{Simpson}}}, \bibinfo {author} {\bibfnamefont {M.}~\bibnamefont {{Snelling}}}, \bibinfo {author} {\bibfnamefont {G.}~\bibnamefont {{Stacy}}}, \bibinfo {author} {\bibfnamefont {H.}~\bibnamefont {{Steinle}}}, \bibinfo {author} {\bibfnamefont {A.}~\bibnamefont {{Strong}}}, \bibinfo {author} {\bibfnamefont {B.~N.}\ \bibnamefont {{Swanenburg}}}, \bibinfo {author} {\bibfnamefont {B.}~\bibnamefont {{Taylor}}}, \bibinfo {author} {\bibfnamefont {C.}~\bibnamefont {{de Vries}}},\ and\ \bibinfo {author} {\bibfnamefont {C.}~\bibnamefont {{Winkler}}},\ }\bibfield  {title} {\bibinfo {title} {{Instrument
  Description and Performance of the Imaging Gamma-Ray Telescope COMPTEL aboard the Compton Gamma-Ray Observatory}},\ }\href {https://doi.org/10.1086/191794} {\bibfield  {journal} {\bibinfo  {journal} {The Astrophysical Journal}\ }\textbf {\bibinfo {volume} {86}},\ \bibinfo {pages} {657} (\bibinfo {year} {1993})}\BibitemShut {NoStop}%
\bibitem [{\citenamefont {{Boggs, S. E.}}\ and\ \citenamefont {{Jean, P.}}(2001)}]{Boggs_2001}%
  \BibitemOpen
  \bibfield  {author} {\bibinfo {author} {\bibnamefont {{Boggs, S. E.}}}\ and\ \bibinfo {author} {\bibnamefont {{Jean, P.}}},\ }\bibfield  {title} {\bibinfo {title} {Performance characteristics of high resolution compton telescopes},\ }\href {https://doi.org/10.1051/0004-6361:20011058} {\bibfield  {journal} {\bibinfo  {journal} {Astronomy and Astrophysics}\ }\textbf {\bibinfo {volume} {376}},\ \bibinfo {pages} {1126} (\bibinfo {year} {2001})}\BibitemShut {NoStop}%
\bibitem [{\citenamefont {{De Angelis}}\ \emph {et~al.}(2018)\citenamefont {{De Angelis}}, \citenamefont {Tatischeff}, \citenamefont {Grenier}, \citenamefont {McEnery}, \citenamefont {Mallamaci}, \citenamefont {Tavani}, \citenamefont {Oberlack}, \citenamefont {Hanlon}, \citenamefont {Walter}, \citenamefont {Argan}, \citenamefont {{Von Ballmoos}}, \citenamefont {Bulgarelli}, \citenamefont {Bykov}, \citenamefont {Hernanz}, \citenamefont {Kanbach}, \citenamefont {Kuvvetli}, \citenamefont {Pearce}, \citenamefont {Zdziarski}, \citenamefont {Conrad}, \citenamefont {Ghisellini}, \citenamefont {Harding}, \citenamefont {Isern}, \citenamefont {Leising}, \citenamefont {Longo}, \citenamefont {Madejski}, \citenamefont {Martinez}, \citenamefont {Mazziotta}, \citenamefont {Paredes}, \citenamefont {Pohl}, \citenamefont {Rando}, \citenamefont {Razzano}, \citenamefont {Aboudan}, \citenamefont {Ackermann}, \citenamefont {Addazi}, \citenamefont {Ajello}, \citenamefont {Albertus}, \citenamefont {Álvarez}, \citenamefont
  {Ambrosi}, \citenamefont {Antón}, \citenamefont {Antonelli}, \citenamefont {Babic}, \citenamefont {Baibussinov}, \citenamefont {Balbo}, \citenamefont {Baldini}, \citenamefont {Balman}, \citenamefont {Bambi}, \citenamefont {{Barres de Almeida}}, \citenamefont {Barrio}, \citenamefont {Bartels}, \citenamefont {Bastieri}, \citenamefont {Bednarek}, \citenamefont {Bernard}, \citenamefont {Bernardini}, \citenamefont {Bernasconi}, \citenamefont {Bertucci}, \citenamefont {Biland}, \citenamefont {Bissaldi}, \citenamefont {Boettcher}, \citenamefont {Bonvicini}, \citenamefont {Bosch-Ramon}, \citenamefont {Bottacini}, \citenamefont {Bozhilov}, \citenamefont {Bretz}, \citenamefont {Branchesi}, \citenamefont {Brdar}, \citenamefont {Bringmann}, \citenamefont {Brogna}, \citenamefont {{Budtz Jørgensen}}, \citenamefont {Busetto}, \citenamefont {Buson}, \citenamefont {Busso}, \citenamefont {Caccianiga}, \citenamefont {Camera}, \citenamefont {Campana}, \citenamefont {Caraveo}, \citenamefont {Cardillo}, \citenamefont
  {Carlson}, \citenamefont {Celestin}, \citenamefont {Cermeño}, \citenamefont {Chen}, \citenamefont {Cheung}, \citenamefont {Churazov}, \citenamefont {Ciprini}, \citenamefont {Coc}, \citenamefont {Colafrancesco}, \citenamefont {Coleiro}, \citenamefont {Collmar}, \citenamefont {Coppi}, \citenamefont {{Curado da Silva}}, \citenamefont {Cutini}, \citenamefont {D'Ammando}, \citenamefont {{De Lotto}}, \citenamefont {{de Martino}}, \citenamefont {{De Rosa}}, \citenamefont {{Del Santo}}, \citenamefont {Delgado}, \citenamefont {Diehl}, \citenamefont {Dietrich}, \citenamefont {Dolgov}, \citenamefont {Domínguez}, \citenamefont {{Dominis Prester}}, \citenamefont {Donnarumma}, \citenamefont {Dorner}, \citenamefont {Doro}, \citenamefont {Dutra}, \citenamefont {Elsaesser}, \citenamefont {Fabrizio}, \citenamefont {Fernández-Barral}, \citenamefont {Fioretti}, \citenamefont {Foffano}, \citenamefont {Formato}, \citenamefont {Fornengo}, \citenamefont {Foschini}, \citenamefont {Franceschini}, \citenamefont {Franckowiak},
  \citenamefont {Funk}, \citenamefont {Fuschino}, \citenamefont {Gaggero}, \citenamefont {Galanti}, \citenamefont {Gargano}, \citenamefont {Gasparrini}, \citenamefont {Gehrz}, \citenamefont {Giammaria}, \citenamefont {Giglietto}, \citenamefont {Giommi}, \citenamefont {Giordano}, \citenamefont {Giroletti}, \citenamefont {Ghirlanda}, \citenamefont {Godinovic}, \citenamefont {Gouiffés}, \citenamefont {Grove}, \citenamefont {Hamadache}, \citenamefont {Hartmann}, \citenamefont {Hayashida}, \citenamefont {Hryczuk}, \citenamefont {Jean}, \citenamefont {Johnson}, \citenamefont {José}, \citenamefont {Kaufmann}, \citenamefont {Khelifi}, \citenamefont {Kiener}, \citenamefont {Knödlseder}, \citenamefont {Kole}, \citenamefont {Kopp}, \citenamefont {Kozhuharov}, \citenamefont {Labanti}, \citenamefont {Lalkovski}, \citenamefont {Laurent}, \citenamefont {Limousin}, \citenamefont {Linares}, \citenamefont {Lindfors}, \citenamefont {Lindner}, \citenamefont {Liu}, \citenamefont {Lombardi}, \citenamefont {Loparco},
  \citenamefont {López-Coto}, \citenamefont {{López Moya}}, \citenamefont {Lott}, \citenamefont {Lubrano}, \citenamefont {Malyshev}, \citenamefont {Mankuzhiyil}, \citenamefont {Mannheim}, \citenamefont {Marchã}, \citenamefont {Marcianò}, \citenamefont {Marcote}, \citenamefont {Mariotti}, \citenamefont {Marisaldi}, \citenamefont {McBreen}, \citenamefont {Mereghetti}, \citenamefont {Merle}, \citenamefont {Mignani}, \citenamefont {Minervini}, \citenamefont {Moiseev}, \citenamefont {Morselli}, \citenamefont {Moura}, \citenamefont {Nakazawa}, \citenamefont {Nava}, \citenamefont {Nieto}, \citenamefont {Orienti}, \citenamefont {Orio}, \citenamefont {Orlando}, \citenamefont {Orleanski}, \citenamefont {Paiano}, \citenamefont {Paoletti}, \citenamefont {Papitto}, \citenamefont {Pasquato}, \citenamefont {Patricelli}, \citenamefont {Pérez-García}, \citenamefont {Persic}, \citenamefont {Piano}, \citenamefont {Pichel}, \citenamefont {Pimenta}, \citenamefont {Pittori}, \citenamefont {Porter}, \citenamefont {Poutanen},
  \citenamefont {Prandini}, \citenamefont {Prantzos}, \citenamefont {Produit}, \citenamefont {Profumo}, \citenamefont {Queiroz}, \citenamefont {Rainó}, \citenamefont {Raklev}, \citenamefont {Regis}, \citenamefont {Reichardt}, \citenamefont {Rephaeli}, \citenamefont {Rico}, \citenamefont {Rodejohann}, \citenamefont {{Rodriguez Fernandez}}, \citenamefont {Roncadelli}, \citenamefont {Roso}, \citenamefont {Rovero}, \citenamefont {Ruffini}, \citenamefont {Sala}, \citenamefont {Sánchez-Conde}, \citenamefont {Santangelo}, \citenamefont {{Saz Parkinson}}, \citenamefont {Sbarrato}, \citenamefont {Shearer}, \citenamefont {Shellard}, \citenamefont {Short}, \citenamefont {Siegert}, \citenamefont {Siqueira}, \citenamefont {Spinelli}, \citenamefont {Stamerra}, \citenamefont {Starrfield}, \citenamefont {Strong}, \citenamefont {Strümke}, \citenamefont {Tavecchio}, \citenamefont {Taverna}, \citenamefont {Terzić}, \citenamefont {Thompson}, \citenamefont {Tibolla}, \citenamefont {Torres}, \citenamefont {Turolla},
  \citenamefont {Ulyanov}, \citenamefont {Ursi}, \citenamefont {Vacchi}, \citenamefont {{Van den Abeele}}, \citenamefont {Vankova-Kirilovai}, \citenamefont {Venter}, \citenamefont {Verrecchia}, \citenamefont {Vincent}, \citenamefont {Wang}, \citenamefont {Weniger}, \citenamefont {Wu}, \citenamefont {Zaharijaš}, \citenamefont {Zampieri}, \citenamefont {Zane}, \citenamefont {Zimmer},\ and\ \citenamefont {Zoglauer}}]{e-ASTROGAM_2018}%
  \BibitemOpen
  \bibfield  {author} {\bibinfo {author} {\bibfnamefont {A.}~\bibnamefont {{De Angelis}}}, \bibinfo {author} {\bibfnamefont {V.}~\bibnamefont {Tatischeff}}, \bibinfo {author} {\bibfnamefont {I.}~\bibnamefont {Grenier}}, \bibinfo {author} {\bibfnamefont {J.}~\bibnamefont {McEnery}}, \bibinfo {author} {\bibfnamefont {M.}~\bibnamefont {Mallamaci}}, \bibinfo {author} {\bibfnamefont {M.}~\bibnamefont {Tavani}}, \bibinfo {author} {\bibfnamefont {U.}~\bibnamefont {Oberlack}}, \bibinfo {author} {\bibfnamefont {L.}~\bibnamefont {Hanlon}}, \bibinfo {author} {\bibfnamefont {R.}~\bibnamefont {Walter}}, \bibinfo {author} {\bibfnamefont {A.}~\bibnamefont {Argan}}, \bibinfo {author} {\bibfnamefont {P.}~\bibnamefont {{Von Ballmoos}}}, \bibinfo {author} {\bibfnamefont {A.}~\bibnamefont {Bulgarelli}}, \bibinfo {author} {\bibfnamefont {A.}~\bibnamefont {Bykov}}, \bibinfo {author} {\bibfnamefont {M.}~\bibnamefont {Hernanz}}, \bibinfo {author} {\bibfnamefont {G.}~\bibnamefont {Kanbach}}, \bibinfo {author} {\bibfnamefont
  {I.}~\bibnamefont {Kuvvetli}}, \bibinfo {author} {\bibfnamefont {M.}~\bibnamefont {Pearce}}, \bibinfo {author} {\bibfnamefont {A.}~\bibnamefont {Zdziarski}}, \bibinfo {author} {\bibfnamefont {J.}~\bibnamefont {Conrad}}, \bibinfo {author} {\bibfnamefont {G.}~\bibnamefont {Ghisellini}}, \bibinfo {author} {\bibfnamefont {A.}~\bibnamefont {Harding}}, \bibinfo {author} {\bibfnamefont {J.}~\bibnamefont {Isern}}, \bibinfo {author} {\bibfnamefont {M.}~\bibnamefont {Leising}}, \bibinfo {author} {\bibfnamefont {F.}~\bibnamefont {Longo}}, \bibinfo {author} {\bibfnamefont {G.}~\bibnamefont {Madejski}}, \bibinfo {author} {\bibfnamefont {M.}~\bibnamefont {Martinez}}, \bibinfo {author} {\bibfnamefont {M.}~\bibnamefont {Mazziotta}}, \bibinfo {author} {\bibfnamefont {J.}~\bibnamefont {Paredes}}, \bibinfo {author} {\bibfnamefont {M.}~\bibnamefont {Pohl}}, \bibinfo {author} {\bibfnamefont {R.}~\bibnamefont {Rando}}, \bibinfo {author} {\bibfnamefont {M.}~\bibnamefont {Razzano}}, \bibinfo {author} {\bibfnamefont
  {A.}~\bibnamefont {Aboudan}}, \bibinfo {author} {\bibfnamefont {M.}~\bibnamefont {Ackermann}}, \bibinfo {author} {\bibfnamefont {A.}~\bibnamefont {Addazi}}, \bibinfo {author} {\bibfnamefont {M.}~\bibnamefont {Ajello}}, \bibinfo {author} {\bibfnamefont {C.}~\bibnamefont {Albertus}}, \bibinfo {author} {\bibfnamefont {J.}~\bibnamefont {Álvarez}}, \bibinfo {author} {\bibfnamefont {G.}~\bibnamefont {Ambrosi}}, \bibinfo {author} {\bibfnamefont {S.}~\bibnamefont {Antón}}, \bibinfo {author} {\bibfnamefont {L.}~\bibnamefont {Antonelli}}, \bibinfo {author} {\bibfnamefont {A.}~\bibnamefont {Babic}}, \bibinfo {author} {\bibfnamefont {B.}~\bibnamefont {Baibussinov}}, \bibinfo {author} {\bibfnamefont {M.}~\bibnamefont {Balbo}}, \bibinfo {author} {\bibfnamefont {L.}~\bibnamefont {Baldini}}, \bibinfo {author} {\bibfnamefont {S.}~\bibnamefont {Balman}}, \bibinfo {author} {\bibfnamefont {C.}~\bibnamefont {Bambi}}, \bibinfo {author} {\bibfnamefont {U.}~\bibnamefont {{Barres de Almeida}}}, \bibinfo {author} {\bibfnamefont
  {J.}~\bibnamefont {Barrio}}, \bibinfo {author} {\bibfnamefont {R.}~\bibnamefont {Bartels}}, \bibinfo {author} {\bibfnamefont {D.}~\bibnamefont {Bastieri}}, \bibinfo {author} {\bibfnamefont {W.}~\bibnamefont {Bednarek}}, \bibinfo {author} {\bibfnamefont {D.}~\bibnamefont {Bernard}}, \bibinfo {author} {\bibfnamefont {E.}~\bibnamefont {Bernardini}}, \bibinfo {author} {\bibfnamefont {T.}~\bibnamefont {Bernasconi}}, \bibinfo {author} {\bibfnamefont {B.}~\bibnamefont {Bertucci}}, \bibinfo {author} {\bibfnamefont {A.}~\bibnamefont {Biland}}, \bibinfo {author} {\bibfnamefont {E.}~\bibnamefont {Bissaldi}}, \bibinfo {author} {\bibfnamefont {M.}~\bibnamefont {Boettcher}}, \bibinfo {author} {\bibfnamefont {V.}~\bibnamefont {Bonvicini}}, \bibinfo {author} {\bibfnamefont {V.}~\bibnamefont {Bosch-Ramon}}, \bibinfo {author} {\bibfnamefont {E.}~\bibnamefont {Bottacini}}, \bibinfo {author} {\bibfnamefont {V.}~\bibnamefont {Bozhilov}}, \bibinfo {author} {\bibfnamefont {T.}~\bibnamefont {Bretz}}, \bibinfo {author}
  {\bibfnamefont {M.}~\bibnamefont {Branchesi}}, \bibinfo {author} {\bibfnamefont {V.}~\bibnamefont {Brdar}}, \bibinfo {author} {\bibfnamefont {T.}~\bibnamefont {Bringmann}}, \bibinfo {author} {\bibfnamefont {A.}~\bibnamefont {Brogna}}, \bibinfo {author} {\bibfnamefont {C.}~\bibnamefont {{Budtz Jørgensen}}}, \bibinfo {author} {\bibfnamefont {G.}~\bibnamefont {Busetto}}, \bibinfo {author} {\bibfnamefont {S.}~\bibnamefont {Buson}}, \bibinfo {author} {\bibfnamefont {M.}~\bibnamefont {Busso}}, \bibinfo {author} {\bibfnamefont {A.}~\bibnamefont {Caccianiga}}, \bibinfo {author} {\bibfnamefont {S.}~\bibnamefont {Camera}}, \bibinfo {author} {\bibfnamefont {R.}~\bibnamefont {Campana}}, \bibinfo {author} {\bibfnamefont {P.}~\bibnamefont {Caraveo}}, \bibinfo {author} {\bibfnamefont {M.}~\bibnamefont {Cardillo}}, \bibinfo {author} {\bibfnamefont {P.}~\bibnamefont {Carlson}}, \bibinfo {author} {\bibfnamefont {S.}~\bibnamefont {Celestin}}, \bibinfo {author} {\bibfnamefont {M.}~\bibnamefont {Cermeño}}, \bibinfo {author}
  {\bibfnamefont {A.}~\bibnamefont {Chen}}, \bibinfo {author} {\bibfnamefont {C.}~\bibnamefont {Cheung}}, \bibinfo {author} {\bibfnamefont {E.}~\bibnamefont {Churazov}}, \bibinfo {author} {\bibfnamefont {S.}~\bibnamefont {Ciprini}}, \bibinfo {author} {\bibfnamefont {A.}~\bibnamefont {Coc}}, \bibinfo {author} {\bibfnamefont {S.}~\bibnamefont {Colafrancesco}}, \bibinfo {author} {\bibfnamefont {A.}~\bibnamefont {Coleiro}}, \bibinfo {author} {\bibfnamefont {W.}~\bibnamefont {Collmar}}, \bibinfo {author} {\bibfnamefont {P.}~\bibnamefont {Coppi}}, \bibinfo {author} {\bibfnamefont {R.}~\bibnamefont {{Curado da Silva}}}, \bibinfo {author} {\bibfnamefont {S.}~\bibnamefont {Cutini}}, \bibinfo {author} {\bibfnamefont {F.}~\bibnamefont {D'Ammando}}, \bibinfo {author} {\bibfnamefont {B.}~\bibnamefont {{De Lotto}}}, \bibinfo {author} {\bibfnamefont {D.}~\bibnamefont {{de Martino}}}, \bibinfo {author} {\bibfnamefont {A.}~\bibnamefont {{De Rosa}}}, \bibinfo {author} {\bibfnamefont {M.}~\bibnamefont {{Del Santo}}}, \bibinfo
  {author} {\bibfnamefont {L.}~\bibnamefont {Delgado}}, \bibinfo {author} {\bibfnamefont {R.}~\bibnamefont {Diehl}}, \bibinfo {author} {\bibfnamefont {S.}~\bibnamefont {Dietrich}}, \bibinfo {author} {\bibfnamefont {A.}~\bibnamefont {Dolgov}}, \bibinfo {author} {\bibfnamefont {A.}~\bibnamefont {Domínguez}}, \bibinfo {author} {\bibfnamefont {D.}~\bibnamefont {{Dominis Prester}}}, \bibinfo {author} {\bibfnamefont {I.}~\bibnamefont {Donnarumma}}, \bibinfo {author} {\bibfnamefont {D.}~\bibnamefont {Dorner}}, \bibinfo {author} {\bibfnamefont {M.}~\bibnamefont {Doro}}, \bibinfo {author} {\bibfnamefont {M.}~\bibnamefont {Dutra}}, \bibinfo {author} {\bibfnamefont {D.}~\bibnamefont {Elsaesser}}, \bibinfo {author} {\bibfnamefont {M.}~\bibnamefont {Fabrizio}}, \bibinfo {author} {\bibfnamefont {A.}~\bibnamefont {Fernández-Barral}}, \bibinfo {author} {\bibfnamefont {V.}~\bibnamefont {Fioretti}}, \bibinfo {author} {\bibfnamefont {L.}~\bibnamefont {Foffano}}, \bibinfo {author} {\bibfnamefont {V.}~\bibnamefont {Formato}},
  \bibinfo {author} {\bibfnamefont {N.}~\bibnamefont {Fornengo}}, \bibinfo {author} {\bibfnamefont {L.}~\bibnamefont {Foschini}}, \bibinfo {author} {\bibfnamefont {A.}~\bibnamefont {Franceschini}}, \bibinfo {author} {\bibfnamefont {A.}~\bibnamefont {Franckowiak}}, \bibinfo {author} {\bibfnamefont {S.}~\bibnamefont {Funk}}, \bibinfo {author} {\bibfnamefont {F.}~\bibnamefont {Fuschino}}, \bibinfo {author} {\bibfnamefont {D.}~\bibnamefont {Gaggero}}, \bibinfo {author} {\bibfnamefont {G.}~\bibnamefont {Galanti}}, \bibinfo {author} {\bibfnamefont {F.}~\bibnamefont {Gargano}}, \bibinfo {author} {\bibfnamefont {D.}~\bibnamefont {Gasparrini}}, \bibinfo {author} {\bibfnamefont {R.}~\bibnamefont {Gehrz}}, \bibinfo {author} {\bibfnamefont {P.}~\bibnamefont {Giammaria}}, \bibinfo {author} {\bibfnamefont {N.}~\bibnamefont {Giglietto}}, \bibinfo {author} {\bibfnamefont {P.}~\bibnamefont {Giommi}}, \bibinfo {author} {\bibfnamefont {F.}~\bibnamefont {Giordano}}, \bibinfo {author} {\bibfnamefont {M.}~\bibnamefont
  {Giroletti}}, \bibinfo {author} {\bibfnamefont {G.}~\bibnamefont {Ghirlanda}}, \bibinfo {author} {\bibfnamefont {N.}~\bibnamefont {Godinovic}}, \bibinfo {author} {\bibfnamefont {C.}~\bibnamefont {Gouiffés}}, \bibinfo {author} {\bibfnamefont {J.}~\bibnamefont {Grove}}, \bibinfo {author} {\bibfnamefont {C.}~\bibnamefont {Hamadache}}, \bibinfo {author} {\bibfnamefont {D.}~\bibnamefont {Hartmann}}, \bibinfo {author} {\bibfnamefont {M.}~\bibnamefont {Hayashida}}, \bibinfo {author} {\bibfnamefont {A.}~\bibnamefont {Hryczuk}}, \bibinfo {author} {\bibfnamefont {P.}~\bibnamefont {Jean}}, \bibinfo {author} {\bibfnamefont {T.}~\bibnamefont {Johnson}}, \bibinfo {author} {\bibfnamefont {J.}~\bibnamefont {José}}, \bibinfo {author} {\bibfnamefont {S.}~\bibnamefont {Kaufmann}}, \bibinfo {author} {\bibfnamefont {B.}~\bibnamefont {Khelifi}}, \bibinfo {author} {\bibfnamefont {J.}~\bibnamefont {Kiener}}, \bibinfo {author} {\bibfnamefont {J.}~\bibnamefont {Knödlseder}}, \bibinfo {author} {\bibfnamefont {M.}~\bibnamefont
  {Kole}}, \bibinfo {author} {\bibfnamefont {J.}~\bibnamefont {Kopp}}, \bibinfo {author} {\bibfnamefont {V.}~\bibnamefont {Kozhuharov}}, \bibinfo {author} {\bibfnamefont {C.}~\bibnamefont {Labanti}}, \bibinfo {author} {\bibfnamefont {S.}~\bibnamefont {Lalkovski}}, \bibinfo {author} {\bibfnamefont {P.}~\bibnamefont {Laurent}}, \bibinfo {author} {\bibfnamefont {O.}~\bibnamefont {Limousin}}, \bibinfo {author} {\bibfnamefont {M.}~\bibnamefont {Linares}}, \bibinfo {author} {\bibfnamefont {E.}~\bibnamefont {Lindfors}}, \bibinfo {author} {\bibfnamefont {M.}~\bibnamefont {Lindner}}, \bibinfo {author} {\bibfnamefont {J.}~\bibnamefont {Liu}}, \bibinfo {author} {\bibfnamefont {S.}~\bibnamefont {Lombardi}}, \bibinfo {author} {\bibfnamefont {F.}~\bibnamefont {Loparco}}, \bibinfo {author} {\bibfnamefont {R.}~\bibnamefont {López-Coto}}, \bibinfo {author} {\bibfnamefont {M.}~\bibnamefont {{López Moya}}}, \bibinfo {author} {\bibfnamefont {B.}~\bibnamefont {Lott}}, \bibinfo {author} {\bibfnamefont {P.}~\bibnamefont
  {Lubrano}}, \bibinfo {author} {\bibfnamefont {D.}~\bibnamefont {Malyshev}}, \bibinfo {author} {\bibfnamefont {N.}~\bibnamefont {Mankuzhiyil}}, \bibinfo {author} {\bibfnamefont {K.}~\bibnamefont {Mannheim}}, \bibinfo {author} {\bibfnamefont {M.}~\bibnamefont {Marchã}}, \bibinfo {author} {\bibfnamefont {A.}~\bibnamefont {Marcianò}}, \bibinfo {author} {\bibfnamefont {B.}~\bibnamefont {Marcote}}, \bibinfo {author} {\bibfnamefont {M.}~\bibnamefont {Mariotti}}, \bibinfo {author} {\bibfnamefont {M.}~\bibnamefont {Marisaldi}}, \bibinfo {author} {\bibfnamefont {S.}~\bibnamefont {McBreen}}, \bibinfo {author} {\bibfnamefont {S.}~\bibnamefont {Mereghetti}}, \bibinfo {author} {\bibfnamefont {A.}~\bibnamefont {Merle}}, \bibinfo {author} {\bibfnamefont {R.}~\bibnamefont {Mignani}}, \bibinfo {author} {\bibfnamefont {G.}~\bibnamefont {Minervini}}, \bibinfo {author} {\bibfnamefont {A.}~\bibnamefont {Moiseev}}, \bibinfo {author} {\bibfnamefont {A.}~\bibnamefont {Morselli}}, \bibinfo {author} {\bibfnamefont {F.}~\bibnamefont
  {Moura}}, \bibinfo {author} {\bibfnamefont {K.}~\bibnamefont {Nakazawa}}, \bibinfo {author} {\bibfnamefont {L.}~\bibnamefont {Nava}}, \bibinfo {author} {\bibfnamefont {D.}~\bibnamefont {Nieto}}, \bibinfo {author} {\bibfnamefont {M.}~\bibnamefont {Orienti}}, \bibinfo {author} {\bibfnamefont {M.}~\bibnamefont {Orio}}, \bibinfo {author} {\bibfnamefont {E.}~\bibnamefont {Orlando}}, \bibinfo {author} {\bibfnamefont {P.}~\bibnamefont {Orleanski}}, \bibinfo {author} {\bibfnamefont {S.}~\bibnamefont {Paiano}}, \bibinfo {author} {\bibfnamefont {R.}~\bibnamefont {Paoletti}}, \bibinfo {author} {\bibfnamefont {A.}~\bibnamefont {Papitto}}, \bibinfo {author} {\bibfnamefont {M.}~\bibnamefont {Pasquato}}, \bibinfo {author} {\bibfnamefont {B.}~\bibnamefont {Patricelli}}, \bibinfo {author} {\bibfnamefont {M.}~\bibnamefont {Pérez-García}}, \bibinfo {author} {\bibfnamefont {M.}~\bibnamefont {Persic}}, \bibinfo {author} {\bibfnamefont {G.}~\bibnamefont {Piano}}, \bibinfo {author} {\bibfnamefont {A.}~\bibnamefont {Pichel}},
  \bibinfo {author} {\bibfnamefont {M.}~\bibnamefont {Pimenta}}, \bibinfo {author} {\bibfnamefont {C.}~\bibnamefont {Pittori}}, \bibinfo {author} {\bibfnamefont {T.}~\bibnamefont {Porter}}, \bibinfo {author} {\bibfnamefont {J.}~\bibnamefont {Poutanen}}, \bibinfo {author} {\bibfnamefont {E.}~\bibnamefont {Prandini}}, \bibinfo {author} {\bibfnamefont {N.}~\bibnamefont {Prantzos}}, \bibinfo {author} {\bibfnamefont {N.}~\bibnamefont {Produit}}, \bibinfo {author} {\bibfnamefont {S.}~\bibnamefont {Profumo}}, \bibinfo {author} {\bibfnamefont {F.}~\bibnamefont {Queiroz}}, \bibinfo {author} {\bibfnamefont {S.}~\bibnamefont {Rainó}}, \bibinfo {author} {\bibfnamefont {A.}~\bibnamefont {Raklev}}, \bibinfo {author} {\bibfnamefont {M.}~\bibnamefont {Regis}}, \bibinfo {author} {\bibfnamefont {I.}~\bibnamefont {Reichardt}}, \bibinfo {author} {\bibfnamefont {Y.}~\bibnamefont {Rephaeli}}, \bibinfo {author} {\bibfnamefont {J.}~\bibnamefont {Rico}}, \bibinfo {author} {\bibfnamefont {W.}~\bibnamefont {Rodejohann}}, \bibinfo
  {author} {\bibfnamefont {G.}~\bibnamefont {{Rodriguez Fernandez}}}, \bibinfo {author} {\bibfnamefont {M.}~\bibnamefont {Roncadelli}}, \bibinfo {author} {\bibfnamefont {L.}~\bibnamefont {Roso}}, \bibinfo {author} {\bibfnamefont {A.}~\bibnamefont {Rovero}}, \bibinfo {author} {\bibfnamefont {R.}~\bibnamefont {Ruffini}}, \bibinfo {author} {\bibfnamefont {G.}~\bibnamefont {Sala}}, \bibinfo {author} {\bibfnamefont {M.}~\bibnamefont {Sánchez-Conde}}, \bibinfo {author} {\bibfnamefont {A.}~\bibnamefont {Santangelo}}, \bibinfo {author} {\bibfnamefont {P.}~\bibnamefont {{Saz Parkinson}}}, \bibinfo {author} {\bibfnamefont {T.}~\bibnamefont {Sbarrato}}, \bibinfo {author} {\bibfnamefont {A.}~\bibnamefont {Shearer}}, \bibinfo {author} {\bibfnamefont {R.}~\bibnamefont {Shellard}}, \bibinfo {author} {\bibfnamefont {K.}~\bibnamefont {Short}}, \bibinfo {author} {\bibfnamefont {T.}~\bibnamefont {Siegert}}, \bibinfo {author} {\bibfnamefont {C.}~\bibnamefont {Siqueira}}, \bibinfo {author} {\bibfnamefont {P.}~\bibnamefont
  {Spinelli}}, \bibinfo {author} {\bibfnamefont {A.}~\bibnamefont {Stamerra}}, \bibinfo {author} {\bibfnamefont {S.}~\bibnamefont {Starrfield}}, \bibinfo {author} {\bibfnamefont {A.}~\bibnamefont {Strong}}, \bibinfo {author} {\bibfnamefont {I.}~\bibnamefont {Strümke}}, \bibinfo {author} {\bibfnamefont {F.}~\bibnamefont {Tavecchio}}, \bibinfo {author} {\bibfnamefont {R.}~\bibnamefont {Taverna}}, \bibinfo {author} {\bibfnamefont {T.}~\bibnamefont {Terzić}}, \bibinfo {author} {\bibfnamefont {D.}~\bibnamefont {Thompson}}, \bibinfo {author} {\bibfnamefont {O.}~\bibnamefont {Tibolla}}, \bibinfo {author} {\bibfnamefont {D.}~\bibnamefont {Torres}}, \bibinfo {author} {\bibfnamefont {R.}~\bibnamefont {Turolla}}, \bibinfo {author} {\bibfnamefont {A.}~\bibnamefont {Ulyanov}}, \bibinfo {author} {\bibfnamefont {A.}~\bibnamefont {Ursi}}, \bibinfo {author} {\bibfnamefont {A.}~\bibnamefont {Vacchi}}, \bibinfo {author} {\bibfnamefont {J.}~\bibnamefont {{Van den Abeele}}}, \bibinfo {author} {\bibfnamefont {G.}~\bibnamefont
  {Vankova-Kirilovai}}, \bibinfo {author} {\bibfnamefont {C.}~\bibnamefont {Venter}}, \bibinfo {author} {\bibfnamefont {F.}~\bibnamefont {Verrecchia}}, \bibinfo {author} {\bibfnamefont {P.}~\bibnamefont {Vincent}}, \bibinfo {author} {\bibfnamefont {X.}~\bibnamefont {Wang}}, \bibinfo {author} {\bibfnamefont {C.}~\bibnamefont {Weniger}}, \bibinfo {author} {\bibfnamefont {X.}~\bibnamefont {Wu}}, \bibinfo {author} {\bibfnamefont {G.}~\bibnamefont {Zaharijaš}}, \bibinfo {author} {\bibfnamefont {L.}~\bibnamefont {Zampieri}}, \bibinfo {author} {\bibfnamefont {S.}~\bibnamefont {Zane}}, \bibinfo {author} {\bibfnamefont {S.}~\bibnamefont {Zimmer}},\ and\ \bibinfo {author} {\bibfnamefont {A.}~\bibnamefont {Zoglauer}},\ }\bibfield  {title} {\bibinfo {title} {Science with e-astrogam: A space mission for mev–gev gamma-ray astrophysics},\ }\href {https://doi.org/https://doi.org/10.1016/j.jheap.2018.07.001} {\bibfield  {journal} {\bibinfo  {journal} {Journal of High Energy Astrophysics}\ }\textbf {\bibinfo {volume}
  {19}},\ \bibinfo {pages} {1} (\bibinfo {year} {2018})}\BibitemShut {NoStop}%
\bibitem [{\citenamefont {{Fleischhack}}(2021)}]{AMEGO_2021}%
  \BibitemOpen
  \bibinfo {editor} {\bibfnamefont {H.}~\bibnamefont {{Fleischhack}}},\ ed.,\ \href@noop {} {\emph {\bibinfo {title} {AMEGO-X: MeV gamma-ray Astronomy in the Multi-messenger Era}}},\ \bibinfo {series} {the 36th International Cosmic Ray Conference}\ No.~\bibinfo {number} {17}\ (\bibinfo {year} {2021})\BibitemShut {NoStop}%
\bibitem [{\citenamefont {Aramaki}\ \emph {et~al.}(2020)\citenamefont {Aramaki}, \citenamefont {Adrian}, \citenamefont {Karagiorgi},\ and\ \citenamefont {Odaka}}]{ARAMAKI_2020}%
  \BibitemOpen
  \bibfield  {author} {\bibinfo {author} {\bibfnamefont {T.}~\bibnamefont {Aramaki}}, \bibinfo {author} {\bibfnamefont {P.~O.~H.}\ \bibnamefont {Adrian}}, \bibinfo {author} {\bibfnamefont {G.}~\bibnamefont {Karagiorgi}},\ and\ \bibinfo {author} {\bibfnamefont {H.}~\bibnamefont {Odaka}},\ }\bibfield  {title} {\bibinfo {title} {Dual mev gamma-ray and dark matter observatory - grams project},\ }\href {https://doi.org/https://doi.org/10.1016/j.astropartphys.2019.07.002} {\bibfield  {journal} {\bibinfo  {journal} {Astroparticle Physics}\ }\textbf {\bibinfo {volume} {114}},\ \bibinfo {pages} {107} (\bibinfo {year} {2020})}\BibitemShut {NoStop}%
\bibitem [{\citenamefont {Tumer}\ \emph {et~al.}(1995)\citenamefont {Tumer}, \citenamefont {Akyuz}, \citenamefont {Bhattacharya}, \citenamefont {Blair}, \citenamefont {Case}, \citenamefont {Dixon}, \citenamefont {Liu}, \citenamefont {O'Neill}, \citenamefont {Samimi}, \citenamefont {White},\ and\ \citenamefont {Zych}}]{TIGER_1995}%
  \BibitemOpen
  \bibfield  {author} {\bibinfo {author} {\bibfnamefont {O.}~\bibnamefont {Tumer}}, \bibinfo {author} {\bibfnamefont {A.}~\bibnamefont {Akyuz}}, \bibinfo {author} {\bibfnamefont {D.}~\bibnamefont {Bhattacharya}}, \bibinfo {author} {\bibfnamefont {S.}~\bibnamefont {Blair}}, \bibinfo {author} {\bibfnamefont {G.}~\bibnamefont {Case}}, \bibinfo {author} {\bibfnamefont {D.}~\bibnamefont {Dixon}}, \bibinfo {author} {\bibfnamefont {C.-J.}\ \bibnamefont {Liu}}, \bibinfo {author} {\bibfnamefont {T.}~\bibnamefont {O'Neill}}, \bibinfo {author} {\bibfnamefont {J.}~\bibnamefont {Samimi}}, \bibinfo {author} {\bibfnamefont {R.}~\bibnamefont {White}},\ and\ \bibinfo {author} {\bibfnamefont {A.}~\bibnamefont {Zych}},\ }\bibfield  {title} {\bibinfo {title} {The tigre instrument for 0.3-100 mev gamma-ray astronomy},\ }\href {https://doi.org/10.1109/23.467770} {\bibfield  {journal} {\bibinfo  {journal} {IEEE Transactions on Nuclear Science}\ }\textbf {\bibinfo {volume} {42}},\ \bibinfo {pages} {907} (\bibinfo {year}
  {1995})}\BibitemShut {NoStop}%
\bibitem [{\citenamefont {Takahashi}\ \emph {et~al.}(2012)\citenamefont {Takahashi}, \citenamefont {Takeda}, \citenamefont {Watanabe},\ and\ \citenamefont {Tajima}}]{Takahashi_2012}%
  \BibitemOpen
  \bibfield  {author} {\bibinfo {author} {\bibfnamefont {T.}~\bibnamefont {Takahashi}}, \bibinfo {author} {\bibfnamefont {S.}~\bibnamefont {Takeda}}, \bibinfo {author} {\bibfnamefont {S.}~\bibnamefont {Watanabe}},\ and\ \bibinfo {author} {\bibfnamefont {H.}~\bibnamefont {Tajima}},\ }\bibfield  {title} {\bibinfo {title} {Visualization of radioactive substances with a si/cdte compton camera},\ }in\ \href {https://doi.org/10.1109/NSSMIC.2012.6551958} {\emph {\bibinfo {booktitle} {2012 IEEE Nuclear Science Symposium and Medical Imaging Conference Record (NSS/MIC)}}}\ (\bibinfo {year} {2012})\ pp.\ \bibinfo {pages} {4199--4204}\BibitemShut {NoStop}%
\bibitem [{\citenamefont {Kabuki}\ \emph {et~al.}(2010)\citenamefont {Kabuki}, \citenamefont {Kimura}, \citenamefont {Amano}, \citenamefont {Nakamoto}, \citenamefont {Kubo}, \citenamefont {Miuchi}, \citenamefont {Kurosawa}, \citenamefont {Takahashi}, \citenamefont {Kawashima}, \citenamefont {Ueda}, \citenamefont {Okada}, \citenamefont {Kubo}, \citenamefont {Kunieda}, \citenamefont {Nakahara}, \citenamefont {Kohara}, \citenamefont {Miyazaki}, \citenamefont {Nakazawa}, \citenamefont {Shirahata}, \citenamefont {Yamamoto}, \citenamefont {Ogawa}, \citenamefont {Togashi}, \citenamefont {Saji},\ and\ \citenamefont {Tanimori}}]{KABUKI2010}%
  \BibitemOpen
  \bibfield  {author} {\bibinfo {author} {\bibfnamefont {S.}~\bibnamefont {Kabuki}}, \bibinfo {author} {\bibfnamefont {H.}~\bibnamefont {Kimura}}, \bibinfo {author} {\bibfnamefont {H.}~\bibnamefont {Amano}}, \bibinfo {author} {\bibfnamefont {Y.}~\bibnamefont {Nakamoto}}, \bibinfo {author} {\bibfnamefont {H.}~\bibnamefont {Kubo}}, \bibinfo {author} {\bibfnamefont {K.}~\bibnamefont {Miuchi}}, \bibinfo {author} {\bibfnamefont {S.}~\bibnamefont {Kurosawa}}, \bibinfo {author} {\bibfnamefont {M.}~\bibnamefont {Takahashi}}, \bibinfo {author} {\bibfnamefont {H.}~\bibnamefont {Kawashima}}, \bibinfo {author} {\bibfnamefont {M.}~\bibnamefont {Ueda}}, \bibinfo {author} {\bibfnamefont {T.}~\bibnamefont {Okada}}, \bibinfo {author} {\bibfnamefont {A.}~\bibnamefont {Kubo}}, \bibinfo {author} {\bibfnamefont {E.}~\bibnamefont {Kunieda}}, \bibinfo {author} {\bibfnamefont {T.}~\bibnamefont {Nakahara}}, \bibinfo {author} {\bibfnamefont {R.}~\bibnamefont {Kohara}}, \bibinfo {author} {\bibfnamefont {O.}~\bibnamefont {Miyazaki}},
  \bibinfo {author} {\bibfnamefont {T.}~\bibnamefont {Nakazawa}}, \bibinfo {author} {\bibfnamefont {T.}~\bibnamefont {Shirahata}}, \bibinfo {author} {\bibfnamefont {E.}~\bibnamefont {Yamamoto}}, \bibinfo {author} {\bibfnamefont {K.}~\bibnamefont {Ogawa}}, \bibinfo {author} {\bibfnamefont {K.}~\bibnamefont {Togashi}}, \bibinfo {author} {\bibfnamefont {H.}~\bibnamefont {Saji}},\ and\ \bibinfo {author} {\bibfnamefont {T.}~\bibnamefont {Tanimori}},\ }\bibfield  {title} {\bibinfo {title} {Electron-tracking compton gamma-ray camera for small animal and phantom imaging},\ }\href {https://doi.org/https://doi.org/10.1016/j.nima.2010.03.085} {\bibfield  {journal} {\bibinfo  {journal} {Nuclear Instruments and Methods in Physics Research Section A: Accelerators, Spectrometers, Detectors and Associated Equipment}\ }\textbf {\bibinfo {volume} {623}},\ \bibinfo {pages} {606} (\bibinfo {year} {2010})},\ \bibinfo {note} {1st International Conference on Technology and Instrumentation in Particle Physics}\BibitemShut {NoStop}%
\bibitem [{\citenamefont {Kataoka}\ \emph {et~al.}(2013)\citenamefont {Kataoka}, \citenamefont {Kishimoto}, \citenamefont {Nishiyama}, \citenamefont {Fujita}, \citenamefont {Takeuchi}, \citenamefont {Kato}, \citenamefont {Nakamori}, \citenamefont {Ohsuka}, \citenamefont {Nakamura}, \citenamefont {Hirayanagi}, \citenamefont {Adachi}, \citenamefont {Uchiyama},\ and\ \citenamefont {Yamamoto}}]{KATAOKA_2013}%
  \BibitemOpen
  \bibfield  {author} {\bibinfo {author} {\bibfnamefont {J.}~\bibnamefont {Kataoka}}, \bibinfo {author} {\bibfnamefont {A.}~\bibnamefont {Kishimoto}}, \bibinfo {author} {\bibfnamefont {T.}~\bibnamefont {Nishiyama}}, \bibinfo {author} {\bibfnamefont {T.}~\bibnamefont {Fujita}}, \bibinfo {author} {\bibfnamefont {K.}~\bibnamefont {Takeuchi}}, \bibinfo {author} {\bibfnamefont {T.}~\bibnamefont {Kato}}, \bibinfo {author} {\bibfnamefont {T.}~\bibnamefont {Nakamori}}, \bibinfo {author} {\bibfnamefont {S.}~\bibnamefont {Ohsuka}}, \bibinfo {author} {\bibfnamefont {S.}~\bibnamefont {Nakamura}}, \bibinfo {author} {\bibfnamefont {M.}~\bibnamefont {Hirayanagi}}, \bibinfo {author} {\bibfnamefont {S.}~\bibnamefont {Adachi}}, \bibinfo {author} {\bibfnamefont {T.}~\bibnamefont {Uchiyama}},\ and\ \bibinfo {author} {\bibfnamefont {K.}~\bibnamefont {Yamamoto}},\ }\bibfield  {title} {\bibinfo {title} {Handy compton camera using 3d position-sensitive scintillators coupled with large-area monolithic mppc arrays},\ }\href
  {https://doi.org/https://doi.org/10.1016/j.nima.2013.07.018} {\bibfield  {journal} {\bibinfo  {journal} {Nuclear Instruments and Methods in Physics Research Section A: Accelerators, Spectrometers, Detectors and Associated Equipment}\ }\textbf {\bibinfo {volume} {732}},\ \bibinfo {pages} {403} (\bibinfo {year} {2013})},\ \bibinfo {note} {vienna Conference on Instrumentation 2013}\BibitemShut {NoStop}%
\bibitem [{\citenamefont {Kuriyama}\ \emph {et~al.}(2022)\citenamefont {Kuriyama}, \citenamefont {Masubuchi}, \citenamefont {Koshikawa}, \citenamefont {Iwashita}, \citenamefont {Omata}, \citenamefont {Kanda}, \citenamefont {Kataoka}, \citenamefont {Tsurumi}, \citenamefont {Diniz}, \citenamefont {Enoto},\ and\ \citenamefont {Wada}}]{Kuriyama_2022}%
  \BibitemOpen
  \bibfield  {author} {\bibinfo {author} {\bibfnamefont {E.}~\bibnamefont {Kuriyama}}, \bibinfo {author} {\bibfnamefont {M.}~\bibnamefont {Masubuchi}}, \bibinfo {author} {\bibfnamefont {N.}~\bibnamefont {Koshikawa}}, \bibinfo {author} {\bibfnamefont {R.}~\bibnamefont {Iwashita}}, \bibinfo {author} {\bibfnamefont {A.}~\bibnamefont {Omata}}, \bibinfo {author} {\bibfnamefont {T.}~\bibnamefont {Kanda}}, \bibinfo {author} {\bibfnamefont {J.}~\bibnamefont {Kataoka}}, \bibinfo {author} {\bibfnamefont {M.}~\bibnamefont {Tsurumi}}, \bibinfo {author} {\bibfnamefont {G.}~\bibnamefont {Diniz}}, \bibinfo {author} {\bibfnamefont {T.}~\bibnamefont {Enoto}},\ and\ \bibinfo {author} {\bibfnamefont {Y.}~\bibnamefont {Wada}},\ }\bibfield  {title} {\bibinfo {title} {Compton camera imaging of a gamma-ray glow from a thunderstorm},\ }\href {https://doi.org/https://doi.org/10.1029/2022GL100139} {\bibfield  {journal} {\bibinfo  {journal} {Geophysical Research Letters}\ }\textbf {\bibinfo {volume} {49}},\ \bibinfo {pages}
  {e2022GL100139} (\bibinfo {year} {2022})},\ \bibinfo {note} {e2022GL100139 2022GL100139}\BibitemShut {NoStop}%
\bibitem [{\citenamefont {Tomono}\ \emph {et~al.}(2017)\citenamefont {Tomono}, \citenamefont {Mizumoto}, \citenamefont {Takada}, \citenamefont {Komura}, \citenamefont {Matsuoka}, \citenamefont {Mizumura}, \citenamefont {Oda},\ and\ \citenamefont {Tanimori}}]{Tomono2017}%
  \BibitemOpen
  \bibfield  {author} {\bibinfo {author} {\bibfnamefont {D.}~\bibnamefont {Tomono}}, \bibinfo {author} {\bibfnamefont {T.}~\bibnamefont {Mizumoto}}, \bibinfo {author} {\bibfnamefont {A.}~\bibnamefont {Takada}}, \bibinfo {author} {\bibfnamefont {S.}~\bibnamefont {Komura}}, \bibinfo {author} {\bibfnamefont {Y.}~\bibnamefont {Matsuoka}}, \bibinfo {author} {\bibfnamefont {Y.}~\bibnamefont {Mizumura}}, \bibinfo {author} {\bibfnamefont {M.}~\bibnamefont {Oda}},\ and\ \bibinfo {author} {\bibfnamefont {T.}~\bibnamefont {Tanimori}},\ }\bibfield  {title} {\bibinfo {title} {{First On-Site True Gamma-Ray Imaging-Spectroscopy of Contamination near Fukushima Plant}},\ }\href {https://doi.org/10.1038/srep41972} {\bibfield  {journal} {\bibinfo  {journal} {Scientific Reports}\ }\textbf {\bibinfo {volume} {7}},\ \bibinfo {pages} {41972} (\bibinfo {year} {2017})}\BibitemShut {NoStop}%
\bibitem [{\citenamefont {Kagaya}\ \emph {et~al.}(2015)\citenamefont {Kagaya}, \citenamefont {Katagiri}, \citenamefont {Enomoto}, \citenamefont {Hanafusa}, \citenamefont {Hosokawa}, \citenamefont {Itoh}, \citenamefont {Muraishi}, \citenamefont {Nakayama}, \citenamefont {Satoh}, \citenamefont {Takeda}, \citenamefont {Tanaka}, \citenamefont {Uchida}, \citenamefont {Watanabe}, \citenamefont {Yanagita}, \citenamefont {Yoshida},\ and\ \citenamefont {Umehara}}]{KAGAYA_2015}%
  \BibitemOpen
  \bibfield  {author} {\bibinfo {author} {\bibfnamefont {M.}~\bibnamefont {Kagaya}}, \bibinfo {author} {\bibfnamefont {H.}~\bibnamefont {Katagiri}}, \bibinfo {author} {\bibfnamefont {R.}~\bibnamefont {Enomoto}}, \bibinfo {author} {\bibfnamefont {R.}~\bibnamefont {Hanafusa}}, \bibinfo {author} {\bibfnamefont {M.}~\bibnamefont {Hosokawa}}, \bibinfo {author} {\bibfnamefont {Y.}~\bibnamefont {Itoh}}, \bibinfo {author} {\bibfnamefont {H.}~\bibnamefont {Muraishi}}, \bibinfo {author} {\bibfnamefont {K.}~\bibnamefont {Nakayama}}, \bibinfo {author} {\bibfnamefont {K.}~\bibnamefont {Satoh}}, \bibinfo {author} {\bibfnamefont {T.}~\bibnamefont {Takeda}}, \bibinfo {author} {\bibfnamefont {M.}~\bibnamefont {Tanaka}}, \bibinfo {author} {\bibfnamefont {T.}~\bibnamefont {Uchida}}, \bibinfo {author} {\bibfnamefont {T.}~\bibnamefont {Watanabe}}, \bibinfo {author} {\bibfnamefont {S.}~\bibnamefont {Yanagita}}, \bibinfo {author} {\bibfnamefont {T.}~\bibnamefont {Yoshida}},\ and\ \bibinfo {author} {\bibfnamefont {K.}~\bibnamefont
  {Umehara}},\ }\bibfield  {title} {\bibinfo {title} {Development of a low-cost-high-sensitivity compton camera using csi (tl) scintillators ($\gamma$i)},\ }\href {https://doi.org/https://doi.org/10.1016/j.nima.2015.09.014} {\bibfield  {journal} {\bibinfo  {journal} {Nuclear Instruments and Methods in Physics Research Section A: Accelerators, Spectrometers, Detectors and Associated Equipment}\ }\textbf {\bibinfo {volume} {804}},\ \bibinfo {pages} {25} (\bibinfo {year} {2015})}\BibitemShut {NoStop}%
\bibitem [{\citenamefont {Kierans}\ \emph {et~al.}(2020)\citenamefont {Kierans}, \citenamefont {Boggs}, \citenamefont {Zoglauer}, \citenamefont {Lowell}, \citenamefont {Sleator}, \citenamefont {Beechert}, \citenamefont {Brandt}, \citenamefont {Jean}, \citenamefont {Lazar}, \citenamefont {Roberts}, \citenamefont {Siegert}, \citenamefont {Tomsick},\ and\ \citenamefont {Ballmoos}}]{Kierans_2020}%
  \BibitemOpen
  \bibfield  {author} {\bibinfo {author} {\bibfnamefont {C.~A.}\ \bibnamefont {Kierans}}, \bibinfo {author} {\bibfnamefont {S.~E.}\ \bibnamefont {Boggs}}, \bibinfo {author} {\bibfnamefont {A.}~\bibnamefont {Zoglauer}}, \bibinfo {author} {\bibfnamefont {A.~W.}\ \bibnamefont {Lowell}}, \bibinfo {author} {\bibfnamefont {C.}~\bibnamefont {Sleator}}, \bibinfo {author} {\bibfnamefont {J.}~\bibnamefont {Beechert}}, \bibinfo {author} {\bibfnamefont {T.~J.}\ \bibnamefont {Brandt}}, \bibinfo {author} {\bibfnamefont {P.}~\bibnamefont {Jean}}, \bibinfo {author} {\bibfnamefont {H.}~\bibnamefont {Lazar}}, \bibinfo {author} {\bibfnamefont {J.}~\bibnamefont {Roberts}}, \bibinfo {author} {\bibfnamefont {T.}~\bibnamefont {Siegert}}, \bibinfo {author} {\bibfnamefont {J.~A.}\ \bibnamefont {Tomsick}},\ and\ \bibinfo {author} {\bibfnamefont {P.~v.}\ \bibnamefont {Ballmoos}},\ }\bibfield  {title} {\bibinfo {title} {Detection of the 511 kev galactic positron annihilation line with cosi},\ }\href
  {https://doi.org/10.3847/1538-4357/ab89a9} {\bibfield  {journal} {\bibinfo  {journal} {The Astrophysical Journal}\ }\textbf {\bibinfo {volume} {895}},\ \bibinfo {pages} {44} (\bibinfo {year} {2020})}\BibitemShut {NoStop}%
\bibitem [{\citenamefont {Strong}(1995)}]{Strong1995}%
  \BibitemOpen
  \bibfield  {author} {\bibinfo {author} {\bibfnamefont {A.~W.}\ \bibnamefont {Strong}},\ }\bibfield  {title} {\bibinfo {title} {{Maximum Entropy imaging of comptel data}},\ }\href {https://doi.org/10.1007/BF00419263} {\bibfield  {journal} {\bibinfo  {journal} {Experimental Astronomy}\ }\textbf {\bibinfo {volume} {6}},\ \bibinfo {pages} {97} (\bibinfo {year} {1995})}\BibitemShut {NoStop}%
\bibitem [{\citenamefont {{Kn{\"o}dlseder}}\ \emph {et~al.}(1999)\citenamefont {{Kn{\"o}dlseder}}, \citenamefont {{Dixon}}, \citenamefont {{Bennett}}, \citenamefont {{Bloemen}}, \citenamefont {{Diehl}}, \citenamefont {{Hermsen}}, \citenamefont {{Oberlack}}, \citenamefont {{Ryan}}, \citenamefont {{Sch{\"o}nfelder}},\ and\ \citenamefont {{von Ballmoos}}}]{Knodlseder_1999}%
  \BibitemOpen
  \bibfield  {author} {\bibinfo {author} {\bibfnamefont {J.}~\bibnamefont {{Kn{\"o}dlseder}}}, \bibinfo {author} {\bibfnamefont {D.}~\bibnamefont {{Dixon}}}, \bibinfo {author} {\bibfnamefont {K.}~\bibnamefont {{Bennett}}}, \bibinfo {author} {\bibfnamefont {H.}~\bibnamefont {{Bloemen}}}, \bibinfo {author} {\bibfnamefont {R.}~\bibnamefont {{Diehl}}}, \bibinfo {author} {\bibfnamefont {W.}~\bibnamefont {{Hermsen}}}, \bibinfo {author} {\bibfnamefont {U.}~\bibnamefont {{Oberlack}}}, \bibinfo {author} {\bibfnamefont {J.}~\bibnamefont {{Ryan}}}, \bibinfo {author} {\bibfnamefont {V.}~\bibnamefont {{Sch{\"o}nfelder}}},\ and\ \bibinfo {author} {\bibfnamefont {P.}~\bibnamefont {{von Ballmoos}}},\ }\bibfield  {title} {\bibinfo {title} {{Image reconstruction of COMPTEL 1.8 MeV (26) AL line data}},\ }\href {https://doi.org/10.48550/arXiv.astro-ph/9903172} {\bibfield  {journal} {\bibinfo  {journal} {Astronomy and Astrophysics}\ }\textbf {\bibinfo {volume} {345}},\ \bibinfo {pages} {813} (\bibinfo {year} {1999})}\BibitemShut
  {NoStop}%
\bibitem [{\citenamefont {Ikeda}\ \emph {et~al.}(2014)\citenamefont {Ikeda}, \citenamefont {Odaka}, \citenamefont {Uemura}, \citenamefont {Takahashi}, \citenamefont {Watanabe},\ and\ \citenamefont {Takeda}}]{IKEDA201446}%
  \BibitemOpen
  \bibfield  {author} {\bibinfo {author} {\bibfnamefont {S.}~\bibnamefont {Ikeda}}, \bibinfo {author} {\bibfnamefont {H.}~\bibnamefont {Odaka}}, \bibinfo {author} {\bibfnamefont {M.}~\bibnamefont {Uemura}}, \bibinfo {author} {\bibfnamefont {T.}~\bibnamefont {Takahashi}}, \bibinfo {author} {\bibfnamefont {S.}~\bibnamefont {Watanabe}},\ and\ \bibinfo {author} {\bibfnamefont {S.}~\bibnamefont {Takeda}},\ }\bibfield  {title} {\bibinfo {title} {Bin mode estimation methods for compton camera imaging},\ }\href {https://doi.org/https://doi.org/10.1016/j.nima.2014.05.081} {\bibfield  {journal} {\bibinfo  {journal} {Nuclear Instruments and Methods in Physics Research Section A: Accelerators, Spectrometers, Detectors and Associated Equipment}\ }\textbf {\bibinfo {volume} {760}},\ \bibinfo {pages} {46} (\bibinfo {year} {2014})}\BibitemShut {NoStop}%
\bibitem [{\citenamefont {Wilderman}\ \emph {et~al.}(1998)\citenamefont {Wilderman}, \citenamefont {Clinthorne}, \citenamefont {Fessler},\ and\ \citenamefont {Rogers}}]{Wilderman_1998}%
  \BibitemOpen
  \bibfield  {author} {\bibinfo {author} {\bibfnamefont {S.}~\bibnamefont {Wilderman}}, \bibinfo {author} {\bibfnamefont {N.}~\bibnamefont {Clinthorne}}, \bibinfo {author} {\bibfnamefont {J.}~\bibnamefont {Fessler}},\ and\ \bibinfo {author} {\bibfnamefont {W.}~\bibnamefont {Rogers}},\ }\bibfield  {title} {\bibinfo {title} {List-mode maximum likelihood reconstruction of compton scatter camera images in nuclear medicine},\ }in\ \href {https://doi.org/10.1109/NSSMIC.1998.773871} {\emph {\bibinfo {booktitle} {1998 IEEE Nuclear Science Symposium Conference Record. 1998 IEEE Nuclear Science Symposium and Medical Imaging Conference (Cat. No.98CH36255)}}},\ Vol.~\bibinfo {volume} {3}\ (\bibinfo {year} {1998})\ pp.\ \bibinfo {pages} {1716--1720 vol.3}\BibitemShut {NoStop}%
\bibitem [{\citenamefont {Kierans}\ \emph {et~al.}(2022)\citenamefont {Kierans}, \citenamefont {Takahashi},\ and\ \citenamefont {Kanbach}}]{Kierans_2022_Springer}%
  \BibitemOpen
  \bibfield  {author} {\bibinfo {author} {\bibfnamefont {C.}~\bibnamefont {Kierans}}, \bibinfo {author} {\bibfnamefont {T.}~\bibnamefont {Takahashi}},\ and\ \bibinfo {author} {\bibfnamefont {G.}~\bibnamefont {Kanbach}},\ }\bibinfo {title} {Compton telescopes for gamma-ray astrophysics},\ in\ \href {https://doi.org/10.1007/978-981-16-4544-0_46-1} {\emph {\bibinfo {booktitle} {Handbook of X-ray and Gamma-ray Astrophysics}}},\ \bibinfo {editor} {edited by\ \bibinfo {editor} {\bibfnamefont {C.}~\bibnamefont {Bambi}}\ and\ \bibinfo {editor} {\bibfnamefont {A.}~\bibnamefont {Santangelo}}}\ (\bibinfo  {publisher} {Springer Nature Singapore},\ \bibinfo {address} {Singapore},\ \bibinfo {year} {2022})\ pp.\ \bibinfo {pages} {1--72}\BibitemShut {NoStop}%
\bibitem [{\citenamefont {Zoglauer}(2005)}]{Zoglauer_thesis_2005}%
  \BibitemOpen
  \bibfield  {author} {\bibinfo {author} {\bibfnamefont {A.~C.}\ \bibnamefont {Zoglauer}},\ }\emph {\bibinfo {title} {First Light for the Next Generation of Compton and Pair Telescopes}},\ \href@noop {} {Ph.D. thesis},\ \bibinfo  {school} {Technische Universität,} (\bibinfo {year} {2005})\BibitemShut {NoStop}%
\bibitem [{\citenamefont {Tanimori}\ \emph {et~al.}(2017)\citenamefont {Tanimori} \emph {et~al.}}]{TANIMORI2017}%
  \BibitemOpen
  \bibfield  {author} {\bibinfo {author} {\bibfnamefont {T.}~\bibnamefont {Tanimori}} \emph {et~al.},\ }\bibfield  {title} {\bibinfo {title} {Establishment of imaging spectroscopy of nuclear gamma-rays based on geometrical optics},\ }\href {https://doi.org/10.1038/srep41511} {\bibfield  {journal} {\bibinfo  {journal} {Scientific Reports}\ }\textbf {\bibinfo {volume} {7}},\ \bibinfo {pages} {41511} (\bibinfo {year} {2017})}\BibitemShut {NoStop}%
\bibitem [{\citenamefont {Schönfelder}(2004)}]{SCHONFELDER2004}%
  \BibitemOpen
  \bibfield  {author} {\bibinfo {author} {\bibfnamefont {V.}~\bibnamefont {Schönfelder}},\ }\bibfield  {title} {\bibinfo {title} {Lessons learnt from comptel for future telescopes},\ }\href {https://doi.org/https://doi.org/10.1016/j.newar.2003.11.027} {\bibfield  {journal} {\bibinfo  {journal} {New Astronomy Reviews}\ }\textbf {\bibinfo {volume} {48}},\ \bibinfo {pages} {193} (\bibinfo {year} {2004})},\ \bibinfo {note} {astronomy with Radioactivities IV and Filling the Sensitivity Gap in MeV Astronomy}\BibitemShut {NoStop}%
\bibitem [{\citenamefont {Ikeda}\ \emph {et~al.}(2021)\citenamefont {Ikeda}, \citenamefont {Takada}, \citenamefont {Abe}, \citenamefont {Yoshikawa}, \citenamefont {Tsuda}, \citenamefont {Ogio}, \citenamefont {Sonoda}, \citenamefont {Mizumura}, \citenamefont {Yoshida},\ and\ \citenamefont {Tanimori}}]{Ikeda2021}%
  \BibitemOpen
  \bibfield  {author} {\bibinfo {author} {\bibfnamefont {T.}~\bibnamefont {Ikeda}}, \bibinfo {author} {\bibfnamefont {A.}~\bibnamefont {Takada}}, \bibinfo {author} {\bibfnamefont {M.}~\bibnamefont {Abe}}, \bibinfo {author} {\bibfnamefont {K.}~\bibnamefont {Yoshikawa}}, \bibinfo {author} {\bibfnamefont {M.}~\bibnamefont {Tsuda}}, \bibinfo {author} {\bibfnamefont {S.}~\bibnamefont {Ogio}}, \bibinfo {author} {\bibfnamefont {S.}~\bibnamefont {Sonoda}}, \bibinfo {author} {\bibfnamefont {Y.}~\bibnamefont {Mizumura}}, \bibinfo {author} {\bibfnamefont {Y.}~\bibnamefont {Yoshida}},\ and\ \bibinfo {author} {\bibfnamefont {T.}~\bibnamefont {Tanimori}},\ }\bibfield  {title} {\bibinfo {title} {Development of convolutional neural networks for an electron-tracking compton camera},\ }\href {https://doi.org/10.1093/ptep/ptab091} {\bibfield  {journal} {\bibinfo  {journal} {Progress of Theoretical and Experimental Physics}\ }\textbf {\bibinfo {volume} {2021}},\ \bibinfo {pages} {083F01} (\bibinfo {year} {2021})}\BibitemShut
  {NoStop}%
\bibitem [{\citenamefont {Tanimori}\ \emph {et~al.}(2015)\citenamefont {Tanimori}, \citenamefont {Kubo}, \citenamefont {Takada}, \citenamefont {Iwaki}, \citenamefont {Komura}, \citenamefont {Kurosawa}, \citenamefont {Matsuoka}, \citenamefont {Miuchi}, \citenamefont {Miyamoto}, \citenamefont {Mizumoto}, \citenamefont {Mizumura}, \citenamefont {Nakamura}, \citenamefont {Nakamura}, \citenamefont {Oda}, \citenamefont {Parker}, \citenamefont {Sawano}, \citenamefont {Sonoda}, \citenamefont {Takemura}, \citenamefont {Tomono},\ and\ \citenamefont {Ueno}}]{Tanimori_2015}%
  \BibitemOpen
  \bibfield  {author} {\bibinfo {author} {\bibfnamefont {T.}~\bibnamefont {Tanimori}}, \bibinfo {author} {\bibfnamefont {H.}~\bibnamefont {Kubo}}, \bibinfo {author} {\bibfnamefont {A.}~\bibnamefont {Takada}}, \bibinfo {author} {\bibfnamefont {S.}~\bibnamefont {Iwaki}}, \bibinfo {author} {\bibfnamefont {S.}~\bibnamefont {Komura}}, \bibinfo {author} {\bibfnamefont {S.}~\bibnamefont {Kurosawa}}, \bibinfo {author} {\bibfnamefont {Y.}~\bibnamefont {Matsuoka}}, \bibinfo {author} {\bibfnamefont {K.}~\bibnamefont {Miuchi}}, \bibinfo {author} {\bibfnamefont {S.}~\bibnamefont {Miyamoto}}, \bibinfo {author} {\bibfnamefont {T.}~\bibnamefont {Mizumoto}}, \bibinfo {author} {\bibfnamefont {Y.}~\bibnamefont {Mizumura}}, \bibinfo {author} {\bibfnamefont {K.}~\bibnamefont {Nakamura}}, \bibinfo {author} {\bibfnamefont {S.}~\bibnamefont {Nakamura}}, \bibinfo {author} {\bibfnamefont {M.}~\bibnamefont {Oda}}, \bibinfo {author} {\bibfnamefont {J.~D.}\ \bibnamefont {Parker}}, \bibinfo {author} {\bibfnamefont {T.}~\bibnamefont
  {Sawano}}, \bibinfo {author} {\bibfnamefont {S.}~\bibnamefont {Sonoda}}, \bibinfo {author} {\bibfnamefont {T.}~\bibnamefont {Takemura}}, \bibinfo {author} {\bibfnamefont {D.}~\bibnamefont {Tomono}},\ and\ \bibinfo {author} {\bibfnamefont {K.}~\bibnamefont {Ueno}},\ }\bibfield  {title} {\bibinfo {title} {An electron-tracking compton telescope for a survey of the deep universe by mev gamma-rays},\ }\href {https://doi.org/10.1088/0004-637X/810/1/28} {\bibfield  {journal} {\bibinfo  {journal} {The Astrophysical Journal}\ }\textbf {\bibinfo {volume} {810}},\ \bibinfo {pages} {28} (\bibinfo {year} {2015})}\BibitemShut {NoStop}%
\bibitem [{\citenamefont {Tanimori}\ \emph {et~al.}(2004)\citenamefont {Tanimori}, \citenamefont {Kubo}, \citenamefont {Miuchi}, \citenamefont {Nagayoshi}, \citenamefont {Orito}, \citenamefont {Takada}, \citenamefont {Takeda},\ and\ \citenamefont {Ueno}}]{TANIMORI2004263}%
  \BibitemOpen
  \bibfield  {author} {\bibinfo {author} {\bibfnamefont {T.}~\bibnamefont {Tanimori}}, \bibinfo {author} {\bibfnamefont {H.}~\bibnamefont {Kubo}}, \bibinfo {author} {\bibfnamefont {K.}~\bibnamefont {Miuchi}}, \bibinfo {author} {\bibfnamefont {T.}~\bibnamefont {Nagayoshi}}, \bibinfo {author} {\bibfnamefont {R.}~\bibnamefont {Orito}}, \bibinfo {author} {\bibfnamefont {A.}~\bibnamefont {Takada}}, \bibinfo {author} {\bibfnamefont {A.}~\bibnamefont {Takeda}},\ and\ \bibinfo {author} {\bibfnamefont {M.}~\bibnamefont {Ueno}},\ }\bibfield  {title} {\bibinfo {title} {Mev $\gamma$-ray imaging detector with micro-tpc},\ }\href {https://doi.org/https://doi.org/10.1016/j.newar.2003.11.038} {\bibfield  {journal} {\bibinfo  {journal} {New Astronomy Reviews}\ }\textbf {\bibinfo {volume} {48}},\ \bibinfo {pages} {263} (\bibinfo {year} {2004})},\ \bibinfo {note} {astronomy with Radioactivities IV and Filling the Sensitivity Gap in MeV Astronomy}\BibitemShut {NoStop}%
\bibitem [{\citenamefont {Ikeda}\ \emph {et~al.}(2023)\citenamefont {Ikeda}, \citenamefont {Takada}, \citenamefont {Takemura}, \citenamefont {Yoshikawa}, \citenamefont {Nakamura}, \citenamefont {Onozaka}, \citenamefont {Abe}, \citenamefont {Tanimori},\ and\ \citenamefont {Mizumura}}]{2023Ikeda}%
  \BibitemOpen
  \bibfield  {author} {\bibinfo {author} {\bibfnamefont {T.}~\bibnamefont {Ikeda}}, \bibinfo {author} {\bibfnamefont {A.}~\bibnamefont {Takada}}, \bibinfo {author} {\bibfnamefont {T.}~\bibnamefont {Takemura}}, \bibinfo {author} {\bibfnamefont {K.}~\bibnamefont {Yoshikawa}}, \bibinfo {author} {\bibfnamefont {Y.}~\bibnamefont {Nakamura}}, \bibinfo {author} {\bibfnamefont {K.}~\bibnamefont {Onozaka}}, \bibinfo {author} {\bibfnamefont {M.}~\bibnamefont {Abe}}, \bibinfo {author} {\bibfnamefont {T.}~\bibnamefont {Tanimori}},\ and\ \bibinfo {author} {\bibfnamefont {Y.}~\bibnamefont {Mizumura}},\ }\bibfield  {title} {\bibinfo {title} {Background contributions in the electron-tracking compton camera aboard smile-$2+$},\ }\href {https://doi.org/10.1103/PhysRevD.108.123013} {\bibfield  {journal} {\bibinfo  {journal} {Phys. Rev. D}\ }\textbf {\bibinfo {volume} {108}},\ \bibinfo {pages} {123013} (\bibinfo {year} {2023})}\BibitemShut {NoStop}%
\end{thebibliography}%

\end{document}